\documentclass[letter,11pt]{article}

\usepackage[english]{babel}
\usepackage[utf8]{inputenc}
\usepackage{amsmath}
\usepackage{amssymb}
\usepackage{amsthm}
\usepackage{bm}
\usepackage{bbm}
\usepackage{cancel}
\usepackage{centernot}
\usepackage{enumerate}
\usepackage{enumitem}
\usepackage{float}
\usepackage{forest}
\usepackage[margin=1in]{geometry}
\usepackage{graphicx}
\usepackage{hyperref}
\usepackage{listings}
\usepackage{mathrsfs}
\usepackage[subtle]{savetrees}
\usepackage{subcaption}
\usepackage{tikz}
\usetikzlibrary{matrix,arrows.meta}
\usetikzlibrary{positioning}
\usepackage{url}

\usepackage[ruled,vlined,linesnumbered]{algorithm2e}

\SetCommentSty{mycommfont}
\SetAlgoVlined

\newtheorem{theorem}{Theorem}[section]
\newtheorem{lemma}[theorem]{Lemma}
\newtheorem{proposition}[theorem]{Proposition}
\newtheorem{corollary}[theorem]{Corollary}
\newtheorem{definition}[theorem]{Definition}

\newtheorem{remark}[theorem]{Remark}

\newcommand{\Pp}{\ensuremath{\mathbb{P}}}
\newcommand{\E}{\ensuremath{\mathbb{E}}}
\newcommand{\R}{\ensuremath{\mathbb{R}}}

\newcommand{\N}{\ensuremath{\mathbb{N}}}

\newcommand{\eps}{\epsilon}

\newcommand{\floor}[1]{\lfloor #1 \rfloor}

\newcommand{\cE}{\mathcal{E}}

\newcommand{\cG}{\mathcal{G}}

\newcommand{\cN}{\mathcal{N}}

\DeclareMathOperator{\depth}{depth}
\newcommand{\Cmax}{\ensuremath{C_{\max}}}

\newcommand{\Ctwo}{\ensuremath{C^{(2)}}}
\newcommand{\Ctwoinfty}{\ensuremath{C^{(2,\infty)}}}
\newcommand{\Ctwolessinfty}{\ensuremath{C^{(2,<\infty)}}}
\newcommand{\Ctwoell}{\ensuremath{C^{(2,\ell)}}}

\newcommand{\Ctwomax}{\ensuremath{C^{(2)}_{\max}}}

\newcommand{\zetatwo}{\ensuremath{\zeta^{(2)}}}

\newcommand{\zetatwolessinfty}{\ensuremath{\zeta^{(2,<\infty)}}}
\newcommand{\zetatwoinfty}{\ensuremath{\zeta^{(2,\infty)}}}

\newcommand{\zetatwoell}{\ensuremath{\zeta^{(2,\ell)}}}
\newcommand{\zetatwolessell}{\ensuremath{\zeta^{(2,<\ell)}}}

\newcommand{\purple}{\textbf{\textcolor{purple}{purple}}}
\newcommand{\red}{\textbf{\textcolor{red}{red}}}
\newcommand{\gray}{\textbf{\textcolor{gray}{gray}}}
\newcommand{\black}{\textbf{black}}

\newcommand{\ds}{\mathsf{ds}}
\newcommand{\us}{\mathsf{us}}

\newcommand{\ius}{\overline{\mathsf{us}}}
\newcommand{\reg}{\mathsf{reg}}
\newcommand{\col}{\mathsf{col}}

\newcommand{\ER}{Erd{\H o}s-R{\' e}nyi} %

\title{Locality via Global Ties:\\Stability of the 2-Core Against Misspecification}
\author{Christian Borgs\;\!\thanks{Department of Electrical Engineering and Computer Sciences, University of California, Berkeley. Email: \texttt{\{borgs,gengzhao\}@berkeley.edu}.}\;\,\thanks{Bakar Institute of Digital Materials for the Planet (BIDMaP), University of California, Berkeley.} \and Geng Zhao\;\!\footnotemark[1]}
\date{\today}

\begin{document}

\maketitle
\thispagestyle{empty}

\begin{abstract}

For many random graph models, the analysis of a related birth process suggests local sampling algorithms for the size of, e.g., the giant connected component, the $k$-core, the size and probability of an epidemic outbreak, etc. In this paper, we study the question of when these algorithms are robust against misspecification of the graph model, for the special case of the 2-core. We show that, for locally converging graphs with bounded average degrees, under a weak notion of expansion, a local sampling algorithm provides robust estimates for the size of both the 2-core and its largest component. Our weak notion of expansion generalizes the classical definition of expansion, while holding for many well-studied random graph models.
 
Our method involves a two-step sprinkling argument. In the first step, we use sprinkling to establish the existence of a non-empty $2$-core inside the giant, while in the second, we use this non-empty $2$-core as seed for a second sprinkling argument to establish that the giant contains a linear sized $2$-core. The second step is based on a novel coloring scheme for the vertices in the tree-part. Our algorithmic results follow from the structural properties for the $2$-core established in the course of our sprinkling arguments.
 
The run-time of our local algorithm is constant independent of the graph size, with the value of the constant depending on the desired asymptotic accuracy $\epsilon$. But given the existential nature of local limits, our arguments do not give any bound on the functional dependence of this constant on $\epsilon$, nor do they give a bound on how large the graph has to be for the asymptotic additive error bound $\epsilon$ to hold.
\end{abstract}
\clearpage
\pagenumbering{arabic}

\newpage

\section{Introduction}

Real world networks are often enormous. For example, the Facebook network involves over 2 billion users as of 2022 \cite{meta2022earnings}, and the internet easily has more than 10 billion nodes \cite{transforma2022iot}.
The sheer sizes of these real world networks makes it hard for an analyst to load the entire graph into memory, or even to fully determine the entire network (e.g., contact tracing over the entire population for the study of epidemics is impossible).
As a result, testing global properties or estimating global quantities can often be hard. Examples of such global quantities include the size of the largest component, the size of the $k$-core, or the probability and size of an epidemic outbreak under a given epidemic model for the spread.

One possible solution to this challenge is to develop local algorithms, that is, algorithms depending only on bounded neighborhoods.
The general framework for local algorithms is to sample a vertex $v$ in the graph $G$, explore its radius $r$ neighborhood (which we denote by $\cN_r(v;G)$), and then compute a local estimate for the desired quantity. This subroutine is repeated for $T$ times and the average is taken as the estimate.
Advantages of local algorithms include reduced memory use, parallelizability for distributed computing settings (each machine doing computation on an independent neighborhood), and scalability (run time not scaling with the network size).

For instance, a natural local algorithm for determining the (relative) size of the largest connected component (often called the ``giant'') in a graph $G$ operates as follows: Repeatedly draw random vertices $v$, and output the fraction of them that are in connected components whose size exceeds a certain threshold (checked, e.g., using breadth first search). Clearly, such an algorithm does not always give approximately correct answers: it has no way to distinguish between a large connected graph $G_0$ and a graph consisting of two copies of $G_0$. However, when $G$ is obtained from some random graph model (e.g., \ER~random graphs, inhomogeneous random graphs, configuration models, etc.) it is know that, with appropriately chosen parameters, the local algorithm gives approximately correct answer with high probability. These models are locally tree-like, and can be approximated with a branching process.

In this paper, we analyze local algorithms for the $2$-core.  Staying a little more general for the moment, the $k$-core $C^{(k)}(G)$ of a graph $G$ is the maximum subgraph of $G$ where all vertices have degree at least $k$.  For many classic random graph models such as the \ER~model \cite{pittel1996sudden}, the configuration model \cite{janson2007simple}, and inhomogeneous random graphs \cite{riordan2008k}, the $2$-core (and more generally, the $k$-core) can be analyzed using a branching process approximation, suggesting a natural local algorithm exploring the neighborhoods of a random vertex.  More precisely, let $\mathcal A^{(k)}(\cN_r(v;G))$ be the event that  $\cN_r(v;G))$ contains a $k$-regular tree of depth $r$ rooted in $v$. A local algorithm would sample $T$ vertices from the input graphs, and then output the fraction of times the event $\mathcal A^{(k)}(\cN_r(v;G))$ occurs as an estimate for the relative size of the $k$-core.
With a little bit of extra work, the results  of   \cite{riordan2008k}  imply that this local algorithm gives indeed a good approximation for the relative size of the $k$-core, provided the input graphs is drawn  from an inhomogeneous random graph model (aka Graphon model).  

Note that in addition to using that these graphs are locally tree-like, the analysis of 
\cite{riordan2008k} relies heavily on the conditional independence in an inhomogeneous random graph model -- once the features are chosen, edges are independent.  But real world networks are neither locally tree-like, nor do they have the independence properties of inhomogeneous random graphs.  This raises the question: 

\vspace{-2pt}
\begin{center}
    \emph{How robust under model misspecification are these local algorithms?}
\end{center}
\vspace{-2pt}

To address this question, we want to find suitable, fairly  general conditions on a growing sequence of graphs $G_n$ such that
these algorithms give approximately correct answers with high probability. In particular, the condition should apply to the random graph models mentioned above. Further, a desirable condition should also generalize to non-tree-like models (e.g., the household models \cite{household_trapman,household_ball,van2016hierarchical}), possibly after modifying the algorithm to allow for a large number of cycles.  

One natural condition to impose is that the graph instance should ``look like'' these models \emph{locally}.  Without specifying a particular model, this idea is naturally captured by the by now standard notion of local convergence \cite{benjamini2011recurrence,Aldous2004}, which requires convergence of the distributions of finite neighborhoods in the sequence.

But such a condition is clearly not enough.  Indeed, running any local algorithm on a  $d$-regular graph with $n$ vertices, versus running it on $\sqrt n$ disjoint copies of $d$ regular size $\sqrt n$ will clearly return essentially identical answers, even though the first graph has a linear sized connected $k$-core (at least once $d$ is large enough), while the second one does not.  
To address this second issue, we will need to add a condition which says that the graph is ``connected enough'' to not fool local algorithms.  A natural such condition would be expansion, but this seems way too strong, since none of the aforementioned models are expanders, even if one restricts oneself to the giant.  

This bring us to the contributions of this paper, which analyzes the special case of the $2$-core.  First, we {\bf formulate a natural generalization of the above local algorithm for the $2$-core} that makes sense for graphs containing many loops, and second we prove that this
algorithm {\bf is robust against misspecifications}; more precisely, it is an asymptotically correct approximation algorithm (see Theorem~\ref{Thm_main_results_robust} below) provided that the sequence of graphs $G_n$ obeys two conditions
\begin{enumerate}[itemsep=-1pt, topsep=3pt]
    \item The sequence is locally convergent;
    \item It is a weak expander sequence of bounded average degree.
\end{enumerate}

Our notion of local convergence is the standard one (and will be reviewed when we state our main theorems), and our  notion of weak expansion is a weakening of the notion of large set expansion introduced in \cite{alimohammadi2021locality}. 
Before precisely defining it, we formulate our algorithm.

To motivate it, we note that if $v$ is in the 2-core of the $r$-neighborhood $\cN_r(v;G)$, then it is clearly in the 2-core of $G$. But even if locally it is not, it could still be in the 2-core of $G$. In fact, for locally tree-like graphs, we can almost never certify a vertex to be in the $2$-core within the local neighborhood. We capture this in the following notation: given $\cN_r(v;G)$, we say that $v$ is potentially in the 2-core if there exists a graph $G'$ such that $\cN_r(v;G)=\mathcal{N}_r(v;G')$ and $v$ is in the 2-core of $G'$ (note that this is easy to check in a breath first search exploration starting from $v$, see appendix for an algorithm that is linear in the number of edges in $\cN_r(v;G)$).  Using this subroutine, we describe the following local algorithm for estimating the size of the $2$-core in a graph, as well as that of its largest component, $\Cmax^{(2)}(G)$, where in this paper, we define the size $|H|$ of a graph $H$ to be its number of vertices.

\begin{algorithm}[H]\label{Alg_two_core}
\caption{Local approximation for the size of $2$-core}
\SetKwInput{KwInput}{Input}                %
\SetKwInput{KwParam}{Param}
\SetKwInput{KwOutput}{Output}              %
\DontPrintSemicolon
\SetAlgoVlined
  \KwInput{Graph $G=(V,E)$, locality parameter $K$, iteration number $T$}
  \KwOutput{Estimated size of $\Ctwo(G)$ and $\Cmax^{(2)}(G)$}
  
\For{$t=1,\ldots,T$}{
    Sample $v\in V$ uniformly at random\;
    Breadth first explore $\cN_{R}(v;G)$ up to the maximal radius $R$ such that $|\cN_{R}(v;G)| \le K$\;
       Set $I^{(2)}_t = 1$ if $v$ is potentially in the $2$-core given the neighborhood $\cN_{R}(v;G)$ and $0$ otherwise\;
    Set $I^{(2,\infty)}_t = 1 $ if $I^{(2)}_t = 1$ and $\cN_{R}(v;G)$ contains a vertex $u$ at distance $R$ from $v$, and 0 otherwise\;
}
\Return{$I^{(2)} := \frac{1}{T}\sum_{t=1}^T I^{(2)}_t$ as the estimated fractional size of $\Ctwo(G)$, 
and $I^{(2,\infty)} := \frac{1}{T}\sum_{t=1}^T I^{(2,\infty)}_t$ as the estimated fractional size of $\Ctwomax(G)$}
\end{algorithm}

\begin{remark}
    (i) The $2$-core candidacy (Line 4-5) can be checked while exploring the neighborhood. See Algorithm~\ref{Alg_two_core_detailed} in Appendix~\ref{Appendix_alg} for an explicit (pseudocode) implementation of the local exploration. (ii) The algorithm can be parallelized by distributing the $T$ independently sampled vertices across different computing nodes, each exploring the neighborhood of one vertex. (iii) In the distributed setting, the time and memory complexity for each computing node is $O(K^2)$ in the worst case, with $K$ typically chosen to be a large constant. Importantly, the complexity does not scale with $|V|$.
\end{remark}

Having formulated our algorithm, we now define our weak notion of expansion.  We use the notion of an $(\epsilon,\delta)$-cut of a graph $G=(V,E)$ from \cite{dhara2017phase}, 
defined as partition of $V$ into two sets $A$ and $A^c$ such that both have size at least $\epsilon |V|$ while the number of edges between $A$ and $A^c$ is at most $\delta |V|$.

\begin{definition}
[Weak expander sequences] We say a possibly random graph sequence $\{G_n\}_{n\in\N}$ is a \emph{weak expander sequence} if for any $\eps\in(0,1/2)$ there exists $\delta > 0$ (depending on $\eps$) such that 
with high probability (whp) $G_n$ has no $(\epsilon,\delta)$-cut.
We say that such a sequence has bounded average degree if there exists a $\bar d<\infty$ such that whp, the average degree of $G_n$ is bounded by $\bar d$.
\end{definition}

As mentioned above, this notion of weak expansion is a slight generalization of the \emph{large-set expansion}, proposed in \cite{alimohammadi2021locality}, with the only difference being that in our definition, $\delta$ can depend on $\epsilon$, while in \cite{alimohammadi2021locality} it cannot.  
As it turns out, this slight modification makes a big difference for the applicability of our results, since in contrast to 
the 
large-set expansion condition, our
condition holds for the giant of typical random graph models, in particular for
the giant components of both the inhomogeneous random graph \cite[Lemma 11.3]{bollobas2007phase} and the configuration model \cite[Proposition 4.4]{dhara2017phase}.

\nobreak

\subsection{Main results}

Our main result states that for 
a sequence of graphs $G_n$, 
Algorithm~\ref{Alg_two_core} gives an asymptotically correct estimate for the relative size of the $2$-core when the sequence is a locally convergent
sequences of weak expanders with bounded average degree.  As it turns out, our result holds for more general sequences of graphs, namely those obtained from the former by edge percolation.

To state these results, we need some notation. 
First, given a locally finite graph $G$ (i.e., a possibly infinite graph with finite degrees) and a vertex $v$ in $G$, %
we use $C(v;G)$ to denote the connected component of $v$ in $G$.
Next, given a graph $G$ and a parameter $p\in[0,1]$, we define a percolated graph, $G(p)$,
by independently deleting each edge with probability $1-p$.  

Finally, we recall that a sequence of graphs
$G_n$ is called {\it locally convergent}, if the distribution of the $r$-neighborhoods of a random vertex in $G_n$ converges for all $r$.  Formally, let 
$\cG^*$ be the space of all rooted, locally finite graphs.  We will identify two rooted graphs in  $\cG^*$ if they are isomorphic, with isomorphism defined by the existence of a bijection between the vertex sets that map the root into the root,  edges into edges, and non-edges into non-edges.  One then defines a distance $d_{loc}$ on $\cG^*$ by setting $d_{loc}(G,G'))=\frac 1{1+R}$ where $R$ is the largest $R$ such that the balls of radius $R$ around the roots in $G$ and  $G'$ are isomorphic.
This metric turns $\cG^*$ into a metric space. 
One then defines a sequence of $G_n$ to be locally convergent if there exists a measure $\mu$ on $\cG^*$  such that for all 
continuous, bounded functions $f$ on $\cG^*$,
the empirical averages
$\frac 1{n}\sum_{v_n\in G_n}f(G_n,v_n)$ converge%
\footnote{If the sequences  $G_n$ is random, this quantity is random, and we require convergence in probability (we will always assume that $\mu$ is deterministic, making the right hand side non-random).}
to $ \E_\mu[f]$, where for notational convenience, we have assume that $G_n$ has precisely $n$ vertices, a convention we use throughout this paper.

Given a sample $(G,o)$ drawn from the limit $\mu$, we then consider the percolated  graph, $G(p)$, and determine the probability, $\zeta(p)$, that the component of the root is infinite,
\begin{equation}\label{zeta}
    \zeta(p)%
    := \Pp_{(G,o)\sim\mu}(|C(o;G(p))|=\infty).
\end{equation}
We call $\zeta$ the percolation function.  Noting that $\zeta$ is non-decreasing, we define the percolation threshold $p_c=p_c(\mu)$ as
    $
    p_c := \inf \{ p \in [0,1] : \zeta(p) > 0 \}
    $.
As we will see in Theorem~\ref{thm:giant} below, for all $p\neq p_c$, the size of the largest component in the percolated sequence $G_n(p)$ divided by $n$ converges to $\zeta(p)$ when the sequence $G_n$ is locally convergent with bounded average degree and obeys our weak expander assumption.  This in particular implies that $p_c$ is a threshold for the appearance of a giant component in $G_n(p)$.

With these preparations, we are able to state our main result.

\begin{theorem}[Local algorithm for the $2$-core]\label{Thm_main_results_robust}
Let $\eps>0$ be some fixed error tolerance. Assume that  $\{G_n\}_{n\in\N}$ is a locally convergent weak expander sequence of bounded average degree, and $p\neq p_c$.  Then there exist constants $K,T$ such that Algorithm~\ref{Alg_two_core} estimates the relative size of the 2-core (of the entire graph and/or of its largest connected component) up to an $\eps$ additive error for the percolated graph $G_n(p)$ with high probability. That is, with input $G_n(p)$, and parameters $K$, $T$ sufficiently large, the output of Algorithm~\ref{Alg_two_core} satisfies
\begin{equation}
    \left|\frac{1}{n}|\Ctwo(G_n(p))| - I^{(2)} \right| \le \eps \quad\text{ and }\quad \left|\frac{1}{n}|\Ctwomax(G_n(p))| - I^{(2,\infty)} \right| \le \eps, %
\end{equation}
 each with probability at least $1-\eps$ for $n$ sufficiently large.  Furthermore, the required number of iterations $T$ can be bounded by $C\frac{1}{\eps^2}\log \frac{1}{\eps}$ for some universal constant $C$.
\end{theorem}

To prove the theorem, we will analyze
the local neighborhoods of a random vertex $v$ in $G_n$, and examine under which condition a random vertex $v$ that is potentially in the $2$-core, given the neighborhood  $\cN_{R}(v;G)$
is actually in the $2$-core. As we will see, asymptotically, as $R\to\infty$, this is the case for most vertices $v$ when the graph is a weak expander of bounded average degree.  As a byproduct of implementing this proof strategy, we will first prove the following theorem about the asymptotic size of $\Ctwo(G(p))$ and $\Cmax^{(2)}(G(p))$.  

To state it, we define the analogue of the percolation function $\zeta$ for the $2$-core.
To this end, we consider an infinite, rooted graph
$(G,o)$ drawn from the limiting distribution $\mu$ of our locally convergent sequence $G_n$, and consider the $2$-core $\Ctwo(G(p))$ of the percolated graph $G(p)$. 
It will be convenient to divide $\Ctwo(G(p))$ into the union of all its finite components, $\Ctwolessinfty(G(p))$, and the remainder, 
$\Ctwoinfty(G(p))=\Ctwo(G(p))\setminus \Ctwolessinfty(G(p))$.
We then define  three quantities: the probability that the root is 
in an infinite component of the $2$-core after percolation, 
\begin{equation}
    \zetatwoinfty(p) := \Pp_{(G,o)\sim\mu}(o \in \Ctwoinfty(G(p))),
\end{equation}
the analogue for finite components, 
 $\zetatwolessinfty(p;\mu) := \Pp_{(G,o)\sim\mu}(o \in \Ctwolessinfty(G(p)))$, and the sum of the two,
 which we can write as
\begin{equation}
    \zetatwo(p) :=\zetatwolessinfty(p) + \zetatwoinfty(p)=
    \Pp_{(G,o)\sim\mu}(o \in \Ctwo(G(p)).
\end{equation}
To get an intuition for this quantity, we note that
if $G_n$ is locally tree-like, then $G$ is almost surely a tree, and $o \in \Ctwo(G(p))$ if and only if there are two disjoint path from $o$ to $\infty$ in $G(p)$.  Note that this will in particular imply that the root lies in an infinite component of the $2$-core of $G(p)$,
i.e.,
$\zetatwo(p) =\zetatwoinfty(p)$.  
But in general, we get contributions from both 
$\zetatwolessinfty(p)$ and $ \zetatwoinfty(p)$.
We call $\zetatwoinfty(p)$
the percolation function for the \emph{giant} $2$-core%
\footnote{It is often more instructive to study the emergence of a \emph{giant} (i.e., global) $2$-core in percolation, as the small components of the $2$-core depend heavily on local structure of the graphs. For instance, consider the following household model, where we replace each vertex in an \ER~random graph with a triangle, attaching existing edges to the three vertices at random independently. For any $p\in(0,1)$, there will always be a constant fraction of the households that are preserved in percolation with probability $p$, yet a giant $2$-core can only exist above the standard percolation threshold.}.

Theorem~\ref{Thm_main_results_robust}   essentially reduces to the following statement on the convergence of relative sizes of $2$-cores to the corresponding percolation functions.

\begin{theorem}[Size of the $2$-core]\label{Thm_main_results_local_two_core}
Let $\{G_n\}_{n\in\N}$ be a sequence of graphs on $n$ vertices converging locally to $\mu$ as $n\to\infty$. Assume that the sequence is a weak expander sequence with bounded average degrees. Then, for all $p\ne p_c$,
\begin{equation}
    \frac{1}{n}|\Ctwo(G_n(p))| \overset{p}{\to} \zetatwo(p) \qquad\text{ and }\qquad \frac{1}{n}|\Ctwomax(G_n(p))| \overset{p}{\to} \zetatwoinfty(p).
\end{equation}
\end{theorem}

\medskip\noindent{\bf Proof ideas.} 
To prove the Theorem~\ref{Thm_main_results_local_two_core} we use a two step sprinkling argument.  In the first step, we use sprinkling to establish the existence of a non-empty $2$-core inside the giant.
This  non-empty $2$-core then serves as the seed for a linear sized $2$-core in a second 
second sprinkling argument.  The second step is based on a 
 novel coloring scheme for the vertices in the tree-part.

\medskip
\noindent{\bf Open directions.} From an algorithmic point of view, it would be interesting to consider what other parameters 
can be well approximated by a local algorithm under our  weak expansion condition, with the size of the  $k$-core for $k\geq 3$ being a natural candidate.  Second, from a more structural perspective, an interesting question is whether the weak expansion property is preserved (for the giant connected component) whp under percolation.

\nobreak

\subsection{Related works}

The notion of a $k$-core (or $k$-degeneracy) was first introduced in \cite{lick1970k,seidman1983network} as a way to capture cohesion and robustness of a network.
The size of the $k$-core was studied for  several random graph models, including the \ER~model \cite{pittel1996sudden}, inhomogeneous random graphs
\cite{riordan2008k}, and the configuration model \cite{janson2007simple}. These models, being locally tree-like, see a $k$-core within a giant connected component when the parameters are above a certain critical threshold. For these models, it is also well known that the giant component is unique in edge percolation, with its size given in terms of the survival probability of a suitable birth process (e.g., \cite{erdHos1960evolution, bollobas2007phase, fountoulakis2007percolation}).

The question whether these results for the giant hold
in some model free settings has only been analyzed recently \cite{alon2004percolation,benjamini2011critical,Krivelevich-RSA2020,sarkar2021note}.
The main assumptions used in these works are local convergence (originally introduced in a different context in \cite{Aldous2004,benjamini2011recurrence}), and some notion of expansion.  Sprinkling arguments are used in several of these papers, even though the technique is of course much older and goes back to Erd\H os and R\'enyi \cite{erdHos1960evolution}.  Note that it is a-priori not clear what should replace the survival probability of a birth process when the graphs are not locally tree-like, and indeed, the results for the {\it size}
of the giant in all these works assume that the local limit is supported on trees. 
The above question was independently answered in \cite{alimohammadi2021locality} and \cite{van2021giant}, where it was realized that the percolation function \eqref{zeta} should take the role of the survival probability when the graph is not locally tree-like.  Continuing in the line of research of
\cite{alon2004percolation,benjamini2011critical,Krivelevich-RSA2020,sarkar2021note}, the authors of \cite{alimohammadi2021locality} then replaced the bounded degree expander assumptions of the previous work
by the weaker assumption of large set expanders of bounded average degree, a condition they then established for many more realistic models, including preferential attachment \cite{price1965networks, barabasi1999emergence}\footnote{Unfortunately, the giant component of an \ER~random graph generally does not satisfy this condition.}.

However, before this paper, very little was known about 
the size of $k$-cores in the absence of  specific modeling assumption.  
Our results generalize the condition in \cite{alimohammadi2021locality}, and determine the size of both the giant and the $2$-core in percolation under this weakened assumption.

Finally, local algorithms on networks have been extensively studied as a solution to the challenge arising from enormous networks. The earliest works date back to \cite{angluin1980local, linial1992locality} from a distributed computing perspective. Lower bounds were established to characterize the tradeoff between locality and approximation quality (e.g., \cite{linial1992locality, naor1993can, kuhn2004cannot, kuhn2006price, ghaffari2017complexity}).
In recent years, there has also been a significant growth of the local computation algorithm community following the work by Rubinfeld et al. \cite{rubinfeld2011fast}, which shares a similar emphasis on locality within massive graphs but employs a different complexity metric.
A variety of problems have been considered, often involving combinatorial optimization on massive networks. Examples include 
maximal independent sets \cite{schneider2008log, ghaffari2022local},
maximal matchings \cite{nguyen2008constant, mansour2013local},
vertex cover \cite{parnas2007approximating, nguyen2008constant},
distance estimation \cite{marko2006distance},
graph spanners \cite{parter2019local, levi2020local},
page rank \cite{andersen2007local},
and cluster detection \cite{radicchi2004defining, macgregor2021local}.

Most closely related to this work is the work of  \cite{alimohammadi2022algorithms}, who consider the problem of local estimation of the probability and size of epidemic outbreaks on networks, giving an algorithm 
with theoretical guarantees for an epidemic model where the duration of an infection is deterministic, provided the underlying contact network is a large set expander of bounded average degree that converges locally.  In spirit our work is  closely  related to this work.

\nobreak
\section{Preliminaries}

\noindent{\bf  Weak expansion and robustness against attacks.}
 We first note that  for any sequence of weak set expanders,
 $\frac 1n|\Cmax(G_n)|\to 1$ in probability.
 Indeed, fix $\epsilon\in (0,1/2)$, and assume that $|\Cmax(G_n)|\leq (1-\epsilon)n$.  By the definition of weak expansion, we can find a $\delta$ such that
 the edge boundary of $\Cmax(G_n)$ is at least $\delta n$ whp.  But the edge boundary of $\Cmax(G_n)$ is empty, giving a contradiction.  
 
The same argument  shows that for any tolerance parameter $\epsilon\in (0,1/2)$,
 there exists a ``safety margin'' $\delta>0$ such that
 whp, no adversarial attacker can
 reduce the size of the giant by more than $\epsilon n$ vertices unless they can delete at least $\delta n$ edges (robustness against adversarial attacks).   It is finally not hard to see that both properties in turn imply weak expansion, showing that weak expansion is equivalent to the two statements that (a) the giant contains all but $o(n)$ vertices, and (b) the giant is robust against adversarial attacks.
 
 Note that this robustness against adversarial attack implies robustness against random edge deletions, i.e., it show that given any $\epsilon\in (0,1/2)$
 there exists $\delta>0$ such
 the size of the giant in the percolated graphs $G_n(p)$ is whp larger than $(1-\epsilon)n$ provided
$p>1-\frac{\delta}{2\bar d}$.

\medskip
\noindent{\bf Size of the giant component in percolation.}
For a slightly stronger notion of expansion, the following theorem was proved in
\cite{alimohammadi2021locality}.
It turns out that the proofs in \cite{alimohammadi2021locality} easily generalize to our notion of expansion, yielding the following theorem.  We sketch its proof in Appendix~\ref{Appendix_giant}.

\begin{theorem}\label{thm:giant}
    Let $\{G_n\}_{n\in\N}$ be a locally converging sequence of weak expanders with bounded average degree, with limit law $\mu\in\cG^*$. Then $\frac{1}{n}|\Cmax(G_n(p))| \overset{p}{\to} \zeta(p)$ for all $p\ne p_c$. 
    Furthermore, $\zeta$ is continuous except possibly at $p_c$.
\end{theorem}

\begin{remark}
    In view of our remarks on weak expansion and robustness against random edge deletions, one easily sees that under the assumptions of the theorem, $\zeta(p)\to 1$ as $p\to 1$, so in particular $p_c<1$.
\end{remark}

\section{Convergence of the relative size of the $2$-core}\label{Sec_two_core_size_conv}

We will now sketch the proof of our main proposition, Proposition~\ref{Prop_twocore_size_lower},  on the convergence of relative size of the $2$-core in $G_n(p)$.
We will focus on $\Ctwomax(G_n(p))$, the giant component of the $2$-core;
the small components (e.g., isolated cycles, etc.) are local and will be handled in Appendix~\ref{appendix_full_two_core} using standard techniques.

\nobreak
\subsection{Emergence of a non-empty $2$-core in the giant above $p_c$}\label{Sec_two_core_exists}

We start with the following proposition, which states that
for $p > p_c$, with positive probability, the root of a graph drawn from the limiting distribution $\mu$ remains in an infinite $2$-core even after percolation.

\begin{proposition}[Emergence of infinite $2$-core above criticality]\label{Prop_emerge_two_core}
    For $\mu$ a local limit of weak expanders, we have $\zetatwoinfty(p) > 0$ whenever $p > p_c$ and $\zetatwoinfty(p) = 0$ whenever $p < p_c$.
\end{proposition}

\begin{proof}[Proof sketch]
To prove the non-trivial part of the proposition (the statement above $p_c$),  we use that by the properties of local limits, namely ergodicity, the probability of the root lying in an infinite component of the $2$-core is non-zero if and only if the probability that there exists a vertex at a finite distance from the root which lies in the infinite $2$-core is non-zero, see Corollary~\ref{Cor_relate_Pr_in_infty_two_core_to_Pr_in_finite_nbrh_of_twocore} in Appendix~\ref{app:infinite-2core}.  To relate this probability to the probability of the an infinite component in the local limit, we will use the following identity,
   \begin{equation}\label{Eqn_proof_lemma_positive_prob_in_nbrh_of_two_core_reduce_to_zeta_p}
        \Pp\bigg(o\in\bigcup_{r\in\N}\cN_r(\Ctwoinfty(G(p)))\bigg) = \Pp(|C(o;G(p))|=\infty),
    \end{equation}
where the $r$-neighborhood $\cN_r(A;G)$ for a vertex set $A\subseteq V$ is simply $\bigcup_{v\in A} \cN_r(v;G)$.  This identity is also proven in Appendix~\ref{app:infinite-2core}, and is related to the fact that for a local limit of weak expanders, it is impossible to have only one path to infinity above $p_c$.
\end{proof}

Given that we already know that above $p_c$ there is a unique giant of linear size (Theorem~\ref{thm:giant}), the second statement of the theorem suggests that percolation on $G_n$ above criticality should also leave a linear sized giant $2$-core, of asymptotic size $n\zetatwoinfty(p)$.  We will prove this in the next subsection, but here we will only establish the following proposition, with the weaker statement that the giant component contains a non-empty $2$-core above $p_c$. 
This non-empty $2$-core will later seed another sprinkling argument to prove the existence of a linear sized $2$-core.

\begin{proposition}[Nontrivial $2$-core of the giant component]\label{Cor_exist_giant_two_core}
    For $p > p_c$, $\Ctwo(\Cmax(G_n(p))) \ne \emptyset$ whp.
\end{proposition}

To sketch the proof of this proposition, we need some notation. 
Recall that $\zetatwoinfty$ is defined in terms of $\Ctwoinfty(G)$, which is the union of the infinite components of the $2$-core of a graph $G$. Let $\cN_\ell^+(v)$ be the graph obtained from $\cN_\ell(v)$
by planting an infinite ray on each vertex 
at 
distance $\ell$ from $v$. We then define 
$\Ctwoell(G)$ as the set of vertices $v$ in $V(G)$ such that $v\in\Ctwoinfty(\cN_\ell^+(v))$.
In other words, $v\in\Ctwoell(G)$ if local information about the $\ell$-neighborhood cannot rule it out as part of the infinite $2$-core.

Since every vertex in $\Ctwoell(G_n(p))$ is in a connected component of size at least $\ell$, and the combined size of all such components except the largest one is arbitrarily small for $\ell$ sufficiently large (this is a consequence of the uniqueness of the giant, or more precisely, of the fact that $|n^{-1}\Cmax(G_n(p))|\to_p \zeta(p)$), we can guarantee $|\Ctwoell(G_n(p))\backslash \Cmax(G_n(p))| \le \eps n$ for any $\eps > 0$ granted that we choose $\ell$ sufficiently large.
The following lower bound on the size of $\Ctwoell(\Cmax(G_n(p)))$ follows immediately, see Appendix~\ref{app:infinite-2core} for the complete argument.

\begin{lemma}\label{Lemma_Ztwoinfty_k_of_C1_at_p_close_to_limit}
For any $\eps > 0$ and $p \in (p_c,1]$, we have for $\ell$ sufficiently large
\begin{equation}
    \frac{1}{n}|\Ctwoell(\Cmax(G_n(p)))| \ge \zetatwoinfty(p) - \eps \qquad\text{ whp}.
\end{equation}
\end{lemma}

Lemma~\ref{Lemma_Ztwoinfty_k_of_C1_at_p_close_to_limit} suggests that about a correct fraction of vertices in the giant locally look like part of the $2$-core.
One might hope to directly argue that most of such vertices are intricately connected and therefore truly in the $2$-core. We don't know how to prove this directly, and therefore first establish Proposition~\ref{Cor_exist_giant_two_core} using a first sprinkling argument, and then use the non-empty $2$-core in the giant as a seed for a second, more complicated sprinkling argument.

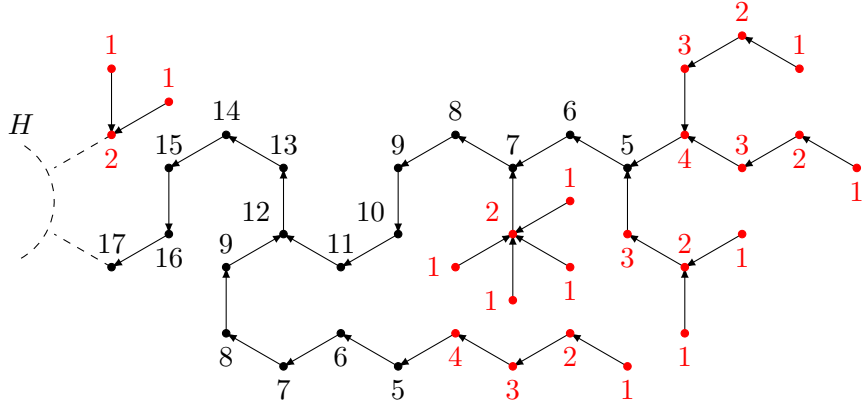
\begin{figure}[H]
  \centering
  
\begin{tikzpicture}[scale=0.44]
\coordinate (H1) at (0,0);
\coordinate (H2) at (0,2);

\coordinate (S17) at ({sqrt(3)},-1);
\coordinate (S16) at ({2*sqrt(3)},0);
\coordinate (S15) at ({2*sqrt(3)},2);
\coordinate (S14) at ({3*sqrt(3)},3);
\coordinate (S13) at ({4*sqrt(3)},2);
\coordinate (S12) at ({4*sqrt(3)},0);
\coordinate (S11) at ({5*sqrt(3)},-1);
\coordinate (S10) at ({6*sqrt(3)},0);
\coordinate (S9) at ({6*sqrt(3)},2);
\coordinate (S8) at ({7*sqrt(3)},3);
\coordinate (S7) at ({8*sqrt(3)},2);
\coordinate (S6) at ({9*sqrt(3)},3);
\coordinate (S5) at ({10*sqrt(3)},2);
\coordinate (S4) at ({11*sqrt(3)},3);
\coordinate (S3) at ({12*sqrt(3)},2);
\coordinate (S2) at ({13*sqrt(3)},3);
\coordinate (S1) at ({14*sqrt(3)},2);

\coordinate (Br1N3) at ({11*sqrt(3)},5);
\coordinate (Br1N2) at ({12*sqrt(3)},6);
\coordinate (Br1N1) at ({13*sqrt(3)},5);

\coordinate (Br2N9) at ({3*sqrt(3)},-1);
\coordinate (Br2N8) at ({3*sqrt(3)},-3);
\coordinate (Br2N7) at ({4*sqrt(3)},-4);
\coordinate (Br2N6) at ({5*sqrt(3)},-3);
\coordinate (Br2N5) at ({6*sqrt(3)},-4);
\coordinate (Br2N4) at ({7*sqrt(3)},-3);
\coordinate (Br2N3) at ({8*sqrt(3)},-4);
\coordinate (Br2N2) at ({9*sqrt(3)},-3);
\coordinate (Br2N1) at ({10*sqrt(3)},-4);

\coordinate (Br3N3) at ({10*sqrt(3)},0);
\coordinate (Br3N2) at ({11*sqrt(3)},-1);
\coordinate (Br3N1) at ({12*sqrt(3)},0);
\coordinate (Br3N1p) at ({11*sqrt(3)},-3);

\coordinate (Br4N2) at ({8*sqrt(3)},0);
\coordinate (Br4N1p1) at ({9*sqrt(3)},1);
\coordinate (Br4N1p2) at ({9*sqrt(3)},-1);
\coordinate (Br4N1p3) at ({8*sqrt(3)},-2);
\coordinate (Br4N1p4) at ({7*sqrt(3)},-1);

\coordinate (Br5N2) at ({sqrt(3)},3);
\coordinate (Br5N1p1) at ({sqrt(3)},5);
\coordinate (Br5N1p2) at ({2*sqrt(3)},4);

\draw[->, >=latex] (S16) -- (S17);
\draw[->, >=latex] (S15) -- (S16);
\draw[->, >=latex] (S14) -- (S15);
\draw[->, >=latex] (S13) -- (S14);
\draw[->, >=latex] (S12) -- (S13);
\draw[->, >=latex] (S11) -- (S12);
\draw[->, >=latex] (S10) -- (S11);
\draw[->, >=latex] (S9) -- (S10);
\draw[->, >=latex] (S8) -- (S9);
\draw[->, >=latex] (S7) -- (S8);
\draw[->, >=latex] (S6) -- (S7);
\draw[->, >=latex] (S5) -- (S6);
\draw[->, >=latex] (S4) -- (S5);
\draw[->, >=latex] (S3) -- (S4);
\draw[->, >=latex] (S2) -- (S3);
\draw[->, >=latex] (S1) -- (S2);

\draw[->, >=latex] (Br1N1) -- (Br1N2);
\draw[->, >=latex] (Br1N2) -- (Br1N3);
\draw[->, >=latex] (Br1N3) -- (S4);

\draw[->, >=latex] (Br2N1) -- (Br2N2);
\draw[->, >=latex] (Br2N2) -- (Br2N3);
\draw[->, >=latex] (Br2N3) -- (Br2N4);
\draw[->, >=latex] (Br2N4) -- (Br2N5);
\draw[->, >=latex] (Br2N5) -- (Br2N6);
\draw[->, >=latex] (Br2N6) -- (Br2N7);
\draw[->, >=latex] (Br2N7) -- (Br2N8);
\draw[->, >=latex] (Br2N8) -- (Br2N9);
\draw[->, >=latex] (Br2N9) -- (S12);

\draw[->, >=latex] (Br3N1) -- (Br3N2);
\draw[->, >=latex] (Br3N1p) -- (Br3N2);
\draw[->, >=latex] (Br3N2) -- (Br3N3);
\draw[->, >=latex] (Br3N3) -- (S5);

\draw[->, >=latex] (Br4N1p1) -- (Br4N2);
\draw[->, >=latex] (Br4N1p2) -- (Br4N2);
\draw[->, >=latex] (Br4N1p3) -- (Br4N2);
\draw[->, >=latex] (Br4N1p4) -- (Br4N2);
\draw[->, >=latex] (Br4N2) -- (S7);

\draw[->, >=latex] (Br5N1p1) -- (Br5N2);
\draw[->, >=latex] (Br5N1p2) -- (Br5N2);

\draw[dashed] (Br5N2) -- (H2);
\draw[dashed] (S17) -- (H1);

\node[circle, draw, fill=black, inner sep=1pt, label=17] at (S17) {};
\node[circle, draw, fill=black, inner sep=1pt, label=below:16] at (S16) {};
\node[circle, draw, fill=black, inner sep=1pt, label=15] at (S15) {};
\node[circle, draw, fill=black, inner sep=1pt, label=14] at (S14) {};
\node[circle, draw, fill=black, inner sep=1pt, label=13] at (S13) {};
\node[circle, draw, fill=black, inner sep=1pt, label=above left:12] at (S12) {};
\node[circle, draw, fill=black, inner sep=1pt, label=11] at (S11) {};
\node[circle, draw, fill=black, inner sep=1pt, label=above left:10] at (S10) {};
\node[circle, draw, fill=black, inner sep=1pt, label=9] at (S9) {};
\node[circle, draw, fill=black, inner sep=1pt, label=8] at (S8) {};
\node[circle, draw, fill=black, inner sep=1pt, label=7] at (S7) {};
\node[circle, draw, fill=black, inner sep=1pt, label=6] at (S6) {};
\node[circle, draw, fill=black, inner sep=1pt, label=5] at (S5) {};
\node[circle, draw=red, fill=red, inner sep=1pt, label={[text=red]below:4}] at (S4) {};
\node[circle, draw=red, fill=red, inner sep=1pt, label={[text=red]3}] at (S3) {};
\node[circle, draw=red, fill=red, inner sep=1pt, label={[text=red]below:2}] at (S2) {};
\node[circle, draw=red, fill=red, inner sep=1pt, label={[text=red]below:1}] at (S1) {};

\node[circle, draw=red, fill=red, inner sep=1pt, label={[text=red]3}] at (Br1N3) {};
\node[circle, draw=red, fill=red, inner sep=1pt, label={[text=red]2}] at (Br1N2) {};
\node[circle, draw=red, fill=red, inner sep=1pt, label={[text=red]1}] at (Br1N1) {};

\node[circle, draw=red, fill=red, inner sep=1pt, label={[text=red]below:1}] at (Br2N1) {};
\node[circle, draw=red, fill=red, inner sep=1pt, label={[text=red]below:2}] at (Br2N2) {};
\node[circle, draw=red, fill=red, inner sep=1pt, label={[text=red]below:3}] at (Br2N3) {};
\node[circle, draw=red, fill=red, inner sep=1pt, label={[text=red]below:4}] at (Br2N4) {};
\node[circle, draw, fill=black, inner sep=1pt, label=below:5] at (Br2N5) {};
\node[circle, draw, fill=black, inner sep=1pt, label=below:6] at (Br2N6) {};
\node[circle, draw, fill=black, inner sep=1pt, label=below:7] at (Br2N7) {};
\node[circle, draw, fill=black, inner sep=1pt, label=below:8] at (Br2N8) {};
\node[circle, draw, fill=black, inner sep=1pt, label=9] at (Br2N9) {};

\node[circle, draw=red, fill=red, inner sep=1pt, label={[text=red]below:3}] at (Br3N3) {};
\node[circle, draw=red, fill=red, inner sep=1pt, label={[text=red]2}] at (Br3N2) {};
\node[circle, draw=red, fill=red, inner sep=1pt, label={[text=red]below:1}] at (Br3N1) {};
\node[circle, draw=red, fill=red, inner sep=1pt, label={[text=red]below:1}] at (Br3N1p) {};

\node[circle, draw=red, fill=red, inner sep=1pt, label={[text=red]above left:2}] at (Br4N2) {};
\node[circle, draw=red, fill=red, inner sep=1pt, label={[text=red]1}] at (Br4N1p1) {};
\node[circle, draw=red, fill=red, inner sep=1pt, label={[text=red]below:1}] at (Br4N1p2) {};
\node[circle, draw=red, fill=red, inner sep=1pt, label={[text=red]left:1}] at (Br4N1p3) {};
\node[circle, draw=red, fill=red, inner sep=1pt, label={[text=red]left:1}] at (Br4N1p4) {};

\node[circle, draw=red, fill=red, inner sep=1pt, label={[text=red]below:2}] at (Br5N2) {};
\node[circle, draw=red, fill=red, inner sep=1pt, label={[text=red]1}] at (Br5N1p1) {};
\node[circle, draw=red, fill=red, inner sep=1pt, label={[text=red]1}] at (Br5N1p2) {};

\draw[dashed] (-1,{1-sqrt(3)}) arc (-60:60:2cm) node[above] {$H$};
\end{tikzpicture}

\caption{The forest $G\backslash H$, oriented in the downstream direction, with vertices annotated with their depths towards $H$. For the choice of $\ell=4$, the red vertices cannot possibly be in $\Ctwoell(G)$.}\label{fig:upstream-downstream}
\end{figure}

\begin{proof}[Proof of Proposition~\ref{Cor_exist_giant_two_core}]
    Fix $p' \in (p_c, p)$. We may assume that $\Cmax(G_n(p'))$ is a tree; otherwise its $2$-core is already nonempty.
    By Lemma~\ref{Lemma_Ztwoinfty_k_of_C1_at_p_close_to_limit}, we know that, for sufficiently large $\ell$, we have 
    \begin{equation}
        \frac{1}{n}|\Ctwoell(\Cmax(G_n(p')))| \ge \eps := \zetatwoinfty(p') / 2 > 0
    \end{equation}
    whp, and we may condition on this event. Consider a connected subgraph $H\subseteq \Cmax(G_n(p'))$ with size $\floor{\eps n / 2}$, constructed, e.g.,  through breadth/depth first search from a leaf of tree. Removing $H$ (i.e., all vertices  along with edges incident to them) from $\Cmax(G_n(p'))$ leaves a forest, i.e., a collection of disjoint trees on vertices in $\Cmax(G_n(p')) \backslash H$.

    At this point, let us define the following terminology, which we will find useful later. Consider a finite \emph{connected} graph $G$ and a nonempty connected subgraph $H\subseteq G$ that contains the $2$-core of $G$ (possibly empty). Then $G\setminus H$ is a forest, where each tree has a unique vertex that connects to $H$ in $G$.  We think of this vertex as its root, and orient the tree towards the root (we call this direction downstream).  We define the
 downstream of $u\notin H$, denoted $\ds(u)$, as the set of vertices strictly between $u$ and $H$, and the upstream of $u$, denoted $\us(u)$, is the set of vertices $v\notin H$ with $u\in\ds(v)$.
 The \emph{depth} of $u\notin H$ towards $H$ in $G$, denoted $\depth_H(u;G)$ is recursively defined as $1$ if $u$ is a leaf node (i.e., has degree $1$) in $G$ and $1+\max_{v\in\us(u)} \depth_H(v;G)$ otherwise.
 Observe that removing $H$ from $G$ leaves a forest. Each tree has a unique vertex with maximal depth $h$, which happens to be the root.
    See Figure~\ref{fig:upstream-downstream} for an illustration.

    Now we return to the setting where $G=\Cmax(G_n(p'))$, with $H\subseteq \Cmax(G_n(p'))$ of size $\floor{\eps n/2}$. 
    Let $F$ be the sub-forest of $\Cmax(G_n(p'))\backslash H$ consisting of trees of size $\ell$ or larger.  {Then} 
    any vertex $v\in\Cmax(G_n(p'))$ not in $H$ or $F$ must have its entire upstream contained in the interior of $\cN_\ell(v;G_n(p'))$; {by the definition of $\Ctwoell$, this implies it cannot be in $\Ctwoell(\Cmax(G_n(p')))$, see again Figure~\ref{fig:upstream-downstream} for an illustration. Put differently, }
    all vertices of $\Ctwoell(\Cmax(G_n(p')))$ must be either in $H$ or $F$.
    Thus,
    \begin{equation}
        |V(F)| \ge |V(F) \cap \Ctwoell(\Cmax(G_n(p')))| \ge |\Ctwoell(\Cmax(G_n(p')))| - |V(H)| \ge \eps n - \frac{\eps n}{2} = \frac{\eps n}{2}.
    \end{equation}
    This leaves us with $H$ and $F$ both containing at least $\eps n/3$ vertices. By weak expansion of $G_n$, the minimum cut between $H$ and $F$ in $G_n$ is at least $\delta n$ for some $\delta$ depending only on $\eps$, which translate through Menger's theorem to $\delta n$ disjoint paths joining $H$ and $F$ in $G_n$. Since $|E(G_n)| \le \bar{d}n/2$ whp, at least $\delta n/2$ of these paths have a length at most $L = \bar{d}/\delta$. On the other hand, $F$ consists of at most 
    $n/\ell$ trees, and thus the number of disjoint paths from $F$ to $H$ in $G_n(p')$ is at most 
    $n/\ell$
    (recall that $\Cmax(G_n(p'))$ is assumed to be a tree). Choose $\ell > 4/\delta$ so that $n/\ell < \delta n / 4$, leaving at least $\delta n / 4$ disjoint paths in $G_n$ that join $H$ with $F$ and are not fully contained in the tree $\Cmax(G_n(p'))$.
    
    We now sprinkle edges with probability $\beta = 1-\frac{1-p}{1-p'}$ to obtain $G_n(p)$ from $G_n(p')$; that is, obtain an independent copy of $G_n(\beta)$ and set $G_n(p)$ to contain the union of (edges in) $G_n(p')$ and $G_n(\beta)$. The probability that none of the $\delta n/4$ disjoint paths of length at most $L$ show up in $G_n(\beta)$ is upper bounded by $(1 - \beta^L)^{\delta n / 4} \to 0 $ as $n\to\infty$.
    Thus there exist, with high probability, two non-identical paths in $\Cmax(G_n(p))$ between some pair of vertices $u\in H$ and $v\in F$, implying that $\Cmax(G_n(p))$ cannot be a tree and must have a nonempty 2-core.
\end{proof}
\nobreak

\subsection{Convergence of the relative size of $2$-core in percolation}\label{Sec_two_core_size_lower}

In this section, we sketch the proof of the convergence in probability  of the relative size of the giant $2$-core of $G_n(p)$, focusing on the lower bound, since the upper bound follows from standard arguments using the local convergence and does not require expansion properties. 
We defer many of the details to the appendices, including the proof of the upper bound for the giant $2$-core (Appendix~\ref{appendix_upper_bound}), and the size of the entire $2$-core (Appendix~\ref{appendix_full_two_core}).
The desired lower bound is given in the following proposition.

\begin{proposition}[Lower bound on the giant 2-core size]\label{Prop_twocore_size_lower}
For any $\eps > 0$ and $p\ne p_c$, we have
\begin{equation}\label{Eqn_prop_twocore_size_lower_main}
    \frac{1}{n}|\Ctwomax(G_n(p))| \ge \frac{1}{n}|\Ctwo(\Cmax(G_n(p)))| \ge \zetatwoinfty(p) - \eps\qquad\text{ whp}.
\end{equation}
\end{proposition}

To prove the proposition, we introduce the following color scheme (see Figure~\ref{fig:coloring} in Appendix~\ref{app:extra figures}
for an illustration).
Let $G=(V,E)$ be a finite connected graph, and let $H\subseteq G$ contain $\Ctwo(G)$. Due to Corollary~\ref{Cor_exist_giant_two_core}, it suffices to consider $\Ctwo(G)\ne \emptyset$. Color vertices and edges of $G$ as follows with a localization parameter $\ell\in\N$:
\begin{enumerate}[topsep=4pt, itemsep=-1pt]
    \item We color a vertex $v\notin H$ \red~if  $\depth_H(v;G)\leq \ell$, and we color it \purple,
    if  $\depth_H(v;G) > 2\ell$ and 
  $(\depth_H(v;G)\;\text{mod}\;\; 2\ell) \in \{1,2,\ldots,\ell\}$; otherwise we leave it \black.
    \item Color each edge between same-colored vertices with \emph{consecutive} depths using the same color as its endpoints. This creates \red~and \purple~connected components in $G\backslash H$, which we call colored \emph{segments}. Call a segment \emph{complete} if it contains a vertex with depth divisible by $\ell$ and \emph{incomplete} otherwise.
    \item In each incomplete \purple~segment, revert the color of the vertices and edges (to \black). In each incomplete \red~segment, color the vertices and edges (along with the edge immediately downstream the segment) \gray, which will be subsequently ignored.
\end{enumerate}
\noindent We will call a vertex outside $H$ \emph{colored} if at the end of this process, it is colored \red~or \purple.  Note that by construction, all colored segments in $G\setminus H$ are trees of height exactly $\ell$, where the maximally upstream vertices have distance exactly $\ell-1$ from the unique maximally downstream vertex; in particular, all segments have size at least $\ell$.

Proposition~\ref{Prop_twocore_size_lower}
will follow from the following two lemmas. To state the first, we assume $p_c<p'<p$. Choose 
$H$ to be
$\Ctwo(\Cmax(G_n(p')))$ if it reaches size $\eps n / 8$; otherwise, expand $\Ctwomax(G_n(p'))$ into a connected subgraph $H\subseteq \Cmax(G_n(p'))$ 
of size $\floor{\frac{\eps n}{8}}$ (in light of Corollary~\ref{Cor_exist_giant_two_core}, we may assume that $\Cmax(G_n(p'))$ has a nontrivial $2$-core). We compute the depths of vertices in $\Cmax(G_n(p'))$ with respect to $H$, and apply the coloring scheme to $\Cmax(G_n(p'))$ with parameter $\ell$, which we will choose later.

\begin{lemma}\label{Lemma_twocore_size_lower_reduction}
Assume $p>p_c$ and $\eps < \zetatwoinfty(p)$. For \eqref{Eqn_prop_twocore_size_lower_main} to hold, it suffices to prove that for $p'<p$ sufficiently close to $p$ and $\ell$ sufficiently large
\begin{equation}\label{Eqn_prop_twocore_size_lower_proof_STP}
    \frac{1}{n} |\Ctwoell(\Cmax(G_n(p'))) \backslash \Ctwo(\Cmax(G_n(p)))| \le \frac{\eps}{2}\qquad\text{ whp}.
\end{equation}
Furthermore, if the event in \eqref{Eqn_prop_twocore_size_lower_proof_STP} fails to happen, then there exist a set of colored segments $S$ in $\Cmax(G_n(p'))$ of combined size at least $\frac{\eps n}{8}$ with $S\cap\Ctwo(G_n(p)) = \emptyset$.
\end{lemma}

The first part of the lemma is straightforward: By continuity of $\zetatwoinfty$ for $p\neq p_c$ (see Appendix~\ref{appendix_continuity}), we may choose $p'\in(p_c,p)$ such that $\zetatwoinfty(p') \ge \zetatwoinfty(p)-\frac{\eps}{4}$. For $\ell$ sufficiently large, the reduction from \eqref{Eqn_prop_twocore_size_lower_main} to \eqref{Eqn_prop_twocore_size_lower_proof_STP} follows from Lemma~\ref{Lemma_Ztwoinfty_k_of_C1_at_p_close_to_limit}.
The proof of the second part involves a detailed analysis of the coloring scheme,
and is 
deferred to Appendix~\ref{app:otherproofs}.

The second lemma will be a consequence of the weak expansion of $G_n$ and its bounded average degree. The proof again uses a sprinkling argument.

\begin{lemma}\label{Lemma_twocore_size_lower_sprinkling}
    There exists $\delta>0$ depending on $\eps$ such that, for $p_c< p' < p$ and for sufficiently large $\ell$, the probability that there exist a set of colored segments $S$ in $\Cmax(G_n(p'))$ of combined size at least $\frac{\eps n}{8}$ with $S\cap \Ctwo(G_n(p)) = \emptyset$ is at most $2^{n/\ell} \cdot (1-\beta^{L})^{\frac{\delta n}{2} - \frac{n}{\ell}}$,
    where $\beta = 1-\frac{1-p}{1-p'}$ and $L = \bar{d}/\delta$. 
\end{lemma}
\begin{proof}
    With $n$ vertices in total and the size of each colored segment at least $\ell$, there are at most $n/\ell$ colored segments and hence at most $2^{n/\ell}$ possible subsets $S$ of them. For each subset $S$ of colored segments with combined size $\frac{\eps n}{8}$ or larger, by weak expansion, there exist at least $\delta n$ \emph{disjoint} paths between these segments and $H$ in the original graph $G_n$, with some $\delta > 0$ depending on $\eps$; further, at least half of these paths have a length at most $L = \bar{d}/\delta$. Let $P$ denote the set of such paths. We obtain $G_n(p)$ from $G_n(p')$ by sprinkling, i.e., recovering each edge in $E(G_n)$ with probability $\beta$ independently, during which each of the paths in $P$ is recovered with probability at least $\beta^L$.

    Focus on a specific colored segment $A$ in $S$, which is a (sub-)tree in $\Cmax(G_n(p'))\backslash H$. Let $u$ be the maximally downstream vertex in $A$, which is unique, and let $v$ be the direct downstream of $u$ (possibly in $H$). If sprinkling recovers a path between $A$ and $H$ that does not go through the direct downstream edge $(u,v)$, then $u$ must be in a cycle and become part of $\Ctwo(G_n(p))$: To see this, note that the removal of $(u,v)$ from $G_n(p)$ leaves $u$ and $v$ still connected -- via the newly recovered path to $H$ and the existing downstream of $v$.

    Now return to the set $P$ of disjoint short paths between $S$ and $H$ in $G_n$. Since $S$ contains at most $n/\ell$ colored segments, at most $n/\ell$ of the paths in $P$ go through a direct downstream edge of a colored segment in $S$.
    For $\ell > \frac{2}{\delta}$, at least $\frac{\delta n}{2} - \frac{n}{\ell}$ disjoint paths in $P$ do not go through the direct downstream edges of the segments in $S$. As is discussed above, if any of these paths is recovered via sprinkling, $S$ cannot be fully outside $\Ctwo(G_n(p))$. %
    Thus, the probability that $S$ is fully outside $\Ctwo(G_n(p))$ is at most $(1-\beta^L)^{\frac{\delta n}{2} - \frac{n}{\ell}}$.
    Taking a union bound over all possible choices of $S$, we ensure that the probability of having a set $S$ of colored segments with combined size at least $\frac{\eps n}{8}$ that remain fully outside $\Ctwo(G_n(p))$ is at most $2^{n/\ell} \cdot (1-\beta^L)^{\frac{\delta n}{2} - \frac{n}{\ell}}$ as we claimed.
\end{proof}

The proof of Proposition~\ref{Prop_twocore_size_lower} immediately follows from the two lemmas, see Appendix~\ref{app:otherproofs} for the details.
Combining Proposition~\ref{Prop_twocore_size_lower} with the  matching upper bound (Lemma~\ref{Lemma_twocore_size_upper} in Appendix~\ref{appendix_upper_bound})
yields the in-probability convergence of the relative size of the giant $2$-core.   The additional analysis needed
to prove in-probability convergence of the relative size of the size of the entire $2$-core will be covered in Appendix~\ref{appendix_full_two_core}.  
Together, these results prove Theorem~\ref{Thm_main_results_local_two_core}.

\section{Robustness of Algorithm~\ref{Alg_two_core}}

Finally, we analyze the robustness of Algorithm~\ref{Alg_two_core} against model misspecification. We offer a sketch of the analysis, with the full proof of Theorem~\ref{Thm_main_results_robust} deferred to Appendix~\ref{Appendix_alg_robust}.

Note that, by %
Theorem~\ref{Thm_main_results_local_two_core},
the actual fractional sizes of $\Ctwo(G_n(p))$ and $\Ctwomax(G_n(p))$ are well approximated by the percolation functions $\zetatwo(p)$ and $\zetatwoinfty(p)$, respectively, for $p\ne p_c$. It then suffices to establish that the estimates $I^{(2)}$ and $I^{(2,\infty)}$ also converge to the $\eps$-neighborhood of the corresponding percolation functions in the limit. For now, let us focus on $I^{(2,\infty)}$ as an estimate of $\zetatwoinfty(p)$; the analysis for $I^{(2)}$ is analogous. In particular, we specify how the parameters can be chosen, from which the proof follows in a straightforward way.

First, choose $\ell$ sufficiently large such that $|\zeta^{(2,\ell')}(p) - \zetatwoinfty(p)| \le \eps/16$ for all $\ell' \ge \ell$. This ensures the quality of the local approximation of the $2$-core whenever $R_t \ge \ell$. The approximation can be poor when $R_t < \ell$, but this only happens when $|\cN_\ell(v_t;G_n)| > K$. By choose $K$ sufficiently large, we can ensure that the probability conditional on $G_n(p)$ (i.e., with respect to the random choice of $v_t$) of $R_t < \ell$ is at most $\eps/8$ whp (with respect to the random $G_n$ and percolation). Overall, the choices of $\ell$ and $K$ guarantees that the expectation of each $I_t^{(2,\infty)}$, conditional on $G_n(p)$, deviates from $\zetatwoinfty(p)$ by at most $\eps/4$ whp. Finally, since $I^{(2,\infty)}$ is the average of $T$ i.i.d. Bernoulli random variables $\{I^{(2,\infty)}_t\}_{t\in[T]}$, it enjoys the classic Hoeffding concentration, and by choosing $T = O\big(\frac{1}{\eps^2} \log \frac{1}{\eps}\big)$, we can ensure that $I^{(2,\infty)}$ deviates from $\E\big[I_t^{(2,\infty)}|G_n(p)\big]$ by at most $\eps/4$ with probability $\eps/2$. With these choices of the parameters, we guarantee that the estimate $I^{(2,\infty)}$ is within an $\frac{\eps}{2}$ additive error from $\zetatwoinfty(p)$.

\newpage

\appendix

\section{Local algorithm for the size of $2$-core}\label{Appendix_alg}

We offer an explicit implementation of Algorithm~\ref{Alg_two_core} in pseudocode. The algorithm consumes a graph $G$, a locality parameter $K$, and the number $T$ of repetitions as inputs, draws $T$ i.i.d. vertices from $V(G)$, runs a local exploration subroutine on each of the samples, and returns the average across the $T$ samples.

Algorithm~\ref{Alg_two_core_detailed} details the local exploration subroutine using breadth-first search. It decides whether a vertex is potentially in the (giant) $2$-core of $G$ based on the local neighborhood.

\begin{algorithm}[H]\label{Alg_two_core_detailed}
\caption{A local exploration subroutine for $2$-core candidacy}
\SetKwInput{KwInput}{Input}                %
\SetKwInput{KwParam}{Param}
\SetKwInput{KwOutput}{Output}              %
\DontPrintSemicolon
\SetAlgoVlined
  \KwInput{Graph $G=(V,E)$, locality parameter $K$}
  \KwOutput{Indicators of potential membership of $\Ctwo(G)$ and $\Ctwo(\Cmax(G))$}
  
    Sample $v\in V$ uniformly at random\;
    \If(\tcp*[f]{a high degree vertex}){$\deg(v) \ge K$}{
        \Return{$I^{(2)}=I^{(2,\infty)}=1$}
    }
    Set $r = 0$, $S = 1$\tcp*{initialize radius of exploration, and size of the neighborhood}
    Initialize $q = \mathsf{Queue}(v)$\;
    \For{$u\in\mathsf{Neighbors}(v)$}{
        Set $r_u = r+1$, $p_u = u$\tcp*{store the distance to $v$ and the ancestor towards $v$}
        Set $\mathsf{reach}_u = r_u$, $\mathsf{cycle}_u = \mathsf{False}$\;
        $S = S+1$\;
        $q.\mathsf{enqueue}(u)$\;
    }
    \While{$q\ne \emptyset$}{
        $u = q.\mathsf{pop}()$\;
        $r = r_u$\;
        \For{$w\in\mathsf{Neighbors}(u)$}{
            \eIf{$p_w = \mathsf{undefined}$}{
                Set $r_w = r+1$, $p_w = p_u$\tcp*{store the distance to $v$ and the ancestor towards $v$}
                Set $\mathsf{reach}_{p_u} = r_w$\;
                $S = S+1$\;
                \If{$S > K$}{
                    \textbf{break for} and \textbf{while}\;
                }
                $q.\mathsf{enqueue}(w)$\;
            }{
                Set $\mathsf{cycle}_{p_u} = \mathsf{True}$\;
            }
        }
    }
\eIf(\tcp*[f]{connected component of size at least $K$}){$S \ge K$}{
    \Return{$I^{(2)} = I^{(2,\infty)} = \mathbbm{1}\big\{\sum_{u\in\mathsf{Neighbors}(v)} \mathbbm{1}\{\mathsf{cycle}_{u} \text{ or }\mathsf{reach}_{u}=r\} \ge 2\big\}$}
}(\tcp*[f]{connected component of size less than $K$}){
\Return{$I^{(2)} = \mathbbm{1}\big\{\sum_{u\in\mathsf{Neighbors}(v)} \mathbbm{1}\{\mathsf{cycle}_{u}\} \ge 2\big\}$ and $I^{(2,\infty)} = 0$}
}
\end{algorithm}

\section{Consequences of local convergence}\label{Appendix_local_conv}

The following fact about percolation on locally converging graph sequences will be used repeated.

\begin{lemma}[Lemma 3.1, \cite{alimohammadi2021locality}]\label{Lemma_3_1_local_event_func_in_prob}
Let $\{G_n\}_{n\in\N}$ be a weak expander sequence converging locally to $(G,o)\in\cG^*$ with law $\mu$. Let $f:\cG^*\to \R$ be a bounded function that depends only on a finite neighborhood around the root. Then, for all $p\in[0,1]$,
\begin{equation}
    \frac{1}{n}\sum_{v\in V(G_n)} f(G_n(p), v) \overset{p}{\to} \E[f(G(p), o)].
\end{equation}
\end{lemma}

The next lemma follows as a corollary.

\begin{lemma}[Bounded neighborhood size]\label{Lemma_bounded_nbrh_size}
    Let $\{G_n\}_{n\in\N}$ be a locally converging sequence with limit $(G,o)\in\cG^*$ with law $\mu$.
    For any $r\in\N$ and $\eps \in(0,1)$, there exists $N \in \N$ such that
    \begin{equation}
        \frac{1}{n} \sum_{v\in V(G_n)} \mathbbm{1}\{|\cN_r(v)| \ge N\} \le \eps \qquad\text{ whp}.
    \end{equation}
\end{lemma}
\begin{proof}
    Let $f_N(G,o) = \mathbbm{1}\{|\cN_r(o)| \ge N\}$, which is clearly local and bounded. By Lemma~\ref{Lemma_3_1_local_event_func_in_prob},
    \begin{equation*}
        \frac{1}{n} \sum_{v\in V(G_n)} f_N(G_n,v) \overset{p}{\to} \E[f_N(G,o)] = \Pp(|\cN_r(o)| \ge N).
    \end{equation*}
    Since $G$ is locally finite a.s., for $N$ sufficiently large, we have $\Pp(|\cN_r(o)| \ge N) \le \eps/2$, which completes the proof.
\end{proof}

\section{Size of the giant component}\label{Appendix_giant}

\subsection{Sprinkling and the size of the giant component}

The proof of Theorem~\ref{thm:giant} is analogous to that of Proposition 3.5 in \cite{alimohammadi2021locality}.
Recall from \cite{alimohammadi2021locality} that we say $G$ is an $(\eps,\alpha)$-large-set expander if
\begin{equation*}
    \phi(G,\eps) := \min_{A\subseteq V, \eps n \le |A| \le (1-\eps)n} \frac{|E(A,V\backslash A)|}{|A|} \ge \alpha.
\end{equation*}
The quantity $\phi(G,\eps)$ is referred to as the $\eps$-large-set expansion of $G$. Our definition of a weak expander sequence simply requires the $\eps$-large-set expansion to be asymptotically bounded away from zero for any $\eps > 0$ (with $\delta = \eps \phi(G,\eps)$ in our definition).

The following lemma restates Lemma~3.3 in \cite{alimohammadi2021locality} and captures the gist of the sprinkling argument. For the ease of reference, we state the proof here, which is identical to the one in \cite{alimohammadi2021locality}.

\begin{lemma}
[Sprinkling. Lemma~3.3, \cite{alimohammadi2021locality}]\label{Lemma_sprinkling} Let $G=(V,E)$ be an $(\eps,\alpha)$-expander on $n$ vertices with average degree at most $\bar d$. Fix real constants $\kappa> 0$ and $\beta,\eps\in(0,1)$. Let $S$ be a collection of disjoint connected subgraphs of $G$ each of size at least $\kappa$. Consider the event that there exists a partition of $S$ into $S_0$ and $S\backslash S_0$ each with size at least $\eps n$ that are disconnected in $G(\beta)$ is at most $\exp\left(\frac{n}{\kappa}- \beta^{\bar{d}/\alpha\eps}\frac{\alpha\eps n}{2}\right)$.
\end{lemma}
\begin{proof}
It suffices to consider $S_0$ consisting of some complete pieces of subgraphs in $S$; otherwise $S_0$ and $S\backslash S_0$ are clearly connected. We hope to lower bound the number of (disjoint) paths joining vertices in $S_0$ and those in $S\backslash S_0$. This is the max-flow between the two subsets of vertices, and is equal to the min-cut. By $\eps$-large-set expansion of $G$, with $|S_0|, |S\backslash S_0| \ge \eps n$, the min-cut is at least $\alpha \eps n$, leading to at least $\alpha\eps n$ disjoint paths between the two sets. Since the total number of edges cannot exceed $n\bar{d}/2$, at least half of the paths have length bounded above by $\frac{\bar{d}}{\alpha\eps}$. Hence, the probability that $S_0$ is disconnected from $S\backslash S_0$ in $G(\beta)$ is at most $(1-\beta^{\bar{d}/\alpha\eps})^{\alpha\eps n/2}$. There are at most $2^{n/\kappa}$ possible choices of $S_0$, and hence the overall probability is bounded above by $2^{n/\kappa} (1-\beta^{\bar{d}/\alpha\eps})^{\alpha\eps n/2} \le \exp(n/\kappa- \beta^{\bar{d}/\alpha\eps}\alpha\eps n/2)$ as desired.
\end{proof}

Note that the sprinkling argument is crafted for any fixed $\eps > 0$. Thus, when proving the uniqueness and locality of the giant component, it suffices to consider a fixed $\eps>0$ and use the $\eps$-large-set expansion (which depends on $\eps$). With this in mind, we consider a locally converging weak expander sequence $\{G_n\}_{n\in\N}$. The exact same analysis as in \cite{alimohammadi2021locality} carries over and suggests the following proposition.

\begin{proposition}\label{Prop_Cmax_converge_at_p_continuity_point}
    For any $\eps > 0$ and any continuity point $p\in[0,1]$ of $\zeta$,
    \begin{equation}\label{Eqn_Cmax_converge_at_p_continuity_point}
        \Pp\Big(\Big|\frac{1}{n}|\Cmax(G_n(p))|-\zeta(p)\Big| \ge \eps\Big) \to 0.
    \end{equation}
\end{proposition}

\subsection{Continuity of the percolation function $\zeta$}
Given Proposition~\ref{Prop_Cmax_converge_at_p_continuity_point}, the convergence of the relative size of $\Cmax(G_n(p))$ as stated in Theorem~\ref{thm:giant} only requires us to establish
the continuity of the percolation function $\zeta$, which, again, nearly follows from the analysis in \cite{alimohammadi2021locality}. The slight generalization of large-set expansion into weak expansion leaves the proof unaffected other than the special attention to pay when we decide the expansion parameter.

First, we observe that the local limit of a weak expander sequence is ergodic.

\begin{lemma}[Cf. Lemma~A.1, \cite{alimohammadi2021locality}]\label{Lemma_ergodicity_of_limit}
    Let $\{G_n\}_{n\in\N}$ be a sequence of (possibly random) weak expanders with bounded average degree converging locally to $(G,o)\in\cG^*$ with law $\mu$. Then $\mu$ is ergodic, and thus extremal among unimodular probability measures on $\cG^*$.
\end{lemma}
\begin{proof}[Proof sketch]
    The proof is almost exactly the same as the proof of Lemma~A.1 in \cite{alimohammadi2021locality}, except for the following minor difference: Once we fix $\eps \le p_0 / 8$, the $\eps$-large-set expansion of our sequence of graphs will be asymptotically $\alpha=\alpha(\eps)$, instead of some absolute quantity (independent of $\eps$) as in the original proof. This, however, causes no issue for us as the rest of the proof relies on the fixed values of $\eps$ and $\alpha$.
\end{proof}

In particular, if $\mu$ is the local limit of a weak expander sequence, then the critical probability of percolation on $(G,o)\sim\mu$ is constant $\mu$-a.s. Let $p_c = p_c(\mu)$ denote this critical probability, i.e.,
\begin{equation}
    \inf \{ p : \Pp(|C(o;G(p))| = \infty |(G,o)) > 0 \} = p_c \qquad \text{a.s.}
\end{equation}
By Theorem~6.7 in \cite{aldous2007processes}, for all $p_1,p_2$ satisfying $p_c < p_1 < p_2\le 1$, every infinite cluster in $G(p_2)$ contains an infinite cluster in $G(p_1)$ (under the standard coupling of percolation) $\mu$-a.s. This allows us to immediately conclude that $\zeta$ is continuous at all $p\ne p_c$.

\begin{proposition}
[Continuity of $\zeta$. Cf. Corollary~2.2, \cite{alimohammadi2021locality}]
\label{Prop_cont_zeta}
Let $\mu$ be the local limit of a weak expander sequence, and $\zeta:[0,1]\to[0,1]$ be the corresponding percolation function. Then $\zeta$ is continuous at all $p\in[0,1]$ except possibly at $p=p_c$.
\end{proposition}

Theorem~\ref{thm:giant} now follows immediately from Propositions~\ref{Prop_Cmax_converge_at_p_continuity_point} and \ref{Prop_cont_zeta}. Further, a slightly more involved analysis suggests that, for all $p\in[0,1]$ (including $p=p_c$), the second largest component $C_2(G_n(p))$ has sublinear size whp (cf. \cite[Appendix~C]{alimohammadi2021locality}).

The following corollary directly follows from the convergence of $|n^{-1}\Cmax(G_n(p))|$ to $\zeta(p)$.
\begin{corollary}\label{Cor_bound_k_large_components_outside_giant}
For any $p\ne p_c$ and any $\eps > 0$, there exists $K$ sufficiently large such that
\begin{equation}
    \frac{1}{n}\sum_{v\in V(G_n)} \mathbbm{1}\{|C(v;G_n(p))| \ge K, v\notin \Cmax(G_n(p))\} \le \eps \qquad\text{ whp}.
\end{equation}
\end{corollary}
\begin{proof}
    The inequality is only interesting when $|\Cmax(G_n(p))| \ge K$ -- otherwise the left hand side is zero.

    When $|\Cmax(G_n(p))| \ge K$, we have $\mathbbm{1}\{|C(v;G_n(p))| \ge K, v\notin \Cmax(G_n(p))\} = \mathbbm{1}\{|C(v;G_n(p))| \ge K\} - \mathbbm{1}\{v\in \Cmax(G_n(p))\}$. Choose $K$ sufficiently large such that $\zeta_{\ge K}(p) := \Pp_{(G,o)\sim\mu}(|C(o;G(p))|\ge K) \le \zeta(p) + \eps/2$, which is possible due to the convergence $\zeta(p) = \lim_{K\to\infty} \zeta_{\ge K}(p)$ (e.g., monotone convergence). By Theorem~\ref{thm:giant},
    \begin{equation*}
        \frac{1}{n}\sum_{v\in V(G_n)} \mathbbm{1}\{v\in \Cmax(G_n(p))\}  = \frac{1}{n}|\Cmax(G_n(p))| \overset{p}{\to} \zeta(p),
    \end{equation*}
    and by Lemma~\ref{Lemma_3_1_local_event_func_in_prob},
    \begin{equation*}
        \frac{1}{n}\sum_{v\in V(G_n)} \mathbbm{1}\{|C(v;G_n(p))| \ge K\} \overset{p}{\to} \zeta_{\ge K}(p).
    \end{equation*}
    Taking the difference between these gives the corollary.
\end{proof}

\section{Continuity of percolation function for the 2-core}\label{appendix_continuity}

We now study the continuity of the functions $\zetatwo$ and $\zetatwoinfty$ for local limits of weak expanders. The proof idea is again very similar to the proof of continuity for the percolation function $\zeta$ in \cite{alimohammadi2021locality} (see Section~2.3 and Appendix~B). For $\zetatwoinfty$, we will analyze the infinite connected components in the percolated graph as is done in \cite{van1984continuity}. The continuity of $\zetatwo$ will be easy to deduce once we have continuity of $\zetatwoinfty$. Throughout this section, we will  be concerned with the random rooted graph $(G,o)\in\cG^*$ with law $\mu$, which is assumed to be the local limit of a weak expander sequence.

\begin{lemma}\label{lem:zeta2inf-cont}
    The function $\zetatwoinfty$, like $\zeta$, is continuous everywhere on $[0,1]$ except possibly at $p_c$.
\end{lemma}
\begin{proof}
Continuity below $p_c$ is trivial since $\zetatwoinfty(p)\equiv 0$ for $p<p_c$. Consider any $p > p_c$. By Proposition~\ref{Prop_emerge_two_core}, $\zetatwoinfty(p) > 0$. By definition,
\begin{equation}
    \zetatwoinfty(p) = \Pp\big(o\in\Ctwoinfty(G(p))\big).
\end{equation}
It is straightforward to verify that $o\in\Ctwoinfty(G(p))$ if and only if $\tilde{H} = C(o;G(p)) \backslash \{o\}$ satisfies one of the following two conditions:
\begin{enumerate}[listparindent=1cm, label=(\Roman*)]
    \item\label{Case_1_two_nbr_to_infty} There exist at least two neighbors $u$ and $v$ of $o$ in $G(p)$ such that $C(u;\tilde{H})$ and $C(v;\tilde{H})$ (possibly the same component) are both infinite;
    \item\label{Case_2_one_nbr_to_infty_and_cycle} There exists exactly one neighbor $u$ of $o$ in $G(p)$ with $|C(u;\tilde{H})|=\infty$ and at least one other neighbor $v$ of $o$ such that $C(o;\tilde{H})$ contains a cycle or contains another neighbor $w\ne u,v$ of $o$.
\end{enumerate}
By Theorem~6.7 in \cite{aldous2007processes}, every infinite connected component of $G(p) \backslash\{o\}$ (and in particular, of $\tilde{H}$) contains an infinite connected component in $G(p')\backslash \{o\}$ for $p' \in (p_c, p)$ a.s.: Otherwise, with a positive probability, we have an infinite component of $G(p)\backslash\{o\}$, connecting to $o$ in $G(p)$ through $k\in\N$ edges, that contains no infinite component of $G(p')$; by resampling the edge percolation on these $k$ edges, we will end up with a positive probability for $G(p)$ to contain an infinite component without an infinite sub-component in $G(p')$, a contradiction with Theorem~6.7 in \cite{aldous2007processes}.

Now it suffices to show that, conditional on $o\in\Ctwoinfty(G(p))$ (this is valid since the probability of such an event is strictly positive), we have $o \in \bigcup_{p''< p} \Ctwoinfty(G(p''))$ a.s. This translates to
\begin{equation}
    \Pp\Bigg(o\in \bigcup_{p''< p} \Ctwoinfty(G(p'')) \Bigg| o \in \Ctwoinfty(G(p))\Bigg) = \sup_{p''<p} \frac{\zetatwoinfty(p'')}{\zetatwoinfty(p)} = 1,
\end{equation}
which immediately suggests left continuity at $p$, and when combined with right continuity of $\zetatwoinfty$ gives our desired result.

To prove the above claim, let $H = C(o;G(p))$, and conditional on $o \in \Ctwoinfty(G(p))$ and the realization of $H$. Note that the cases \ref{Case_1_two_nbr_to_infty} and \ref{Case_2_one_nbr_to_infty_and_cycle} are mutually exclusive. First consider case \ref{Case_1_two_nbr_to_infty}. Choose $p' \in (p_c, p)$. We may assume that $C(u;\tilde{H})$ and $C(v;\tilde{H})$ each contains an infinite connected component of $G(p')$, and hence there exist finite paths from $u,v$ to them in $G(p)$. By sending $p''<p$ sufficiently close to $p$, these paths along with the edges $(o,u)$ and $(o,v)$ will ultimately be included in $G(p'')$, which suggests $o\in\Ctwoinfty(G(p'')))$. For case \ref{Case_2_one_nbr_to_infty_and_cycle}, the reasoning for $u$ remains unchanged, and there is a finite path from $v$ to a (finite) cycle or to $w$ in $G(p)$. This path will again be preserved in $G(p'')$ for $p''<p$ sufficiently close to $p$, suggesting $o\in\Ctwoinfty(G(p'')))$. The proves the claim and completes our proof.
\end{proof}

\begin{corollary}
    The function $\zetatwo$, like $\zeta$, is continuous everywhere on $[0,1]$ except possibly at $p_c$.
\end{corollary}

\begin{proof}
    To study continuity of $\zetatwo$, we consider its point-wise difference with $\zetatwoinfty$, denoted as $\zetatwolessinfty(p) := \zetatwo(p) - \zetatwoinfty(p) = \Pp(o\in\Ctwoinfty(G(p)), |C(o;G(p))| < \infty)$. Since we are not in a finite connected component, it should be clear that, conditional on the realization of $C(o;G(p))$, for $p''<p$ sufficiently close to $p$, no edges in $C(o;G(p))$ are lost and hence $o\in\Ctwolessinfty(G(p''))$ with $|C(o;G(p''))| < \infty$. Thus, $\zetatwolessinfty$ is left continuous on the entire unit interval. Right continuity should also be obvious. Hence, $\zetatwolessinfty$ is continuous, and $\zetatwo = \zetatwolessinfty + \zetatwoinfty$ is continuous on $[0,1]$ except possibly at $p_c$.
\end{proof}

\section{Upper bound}\label{appendix_upper_bound}

In this part, we prove a high-probability upper bound on the relative size of $\Ctwo(\Cmax(G_n(p)))$ using standard techniques (not relying on any expansion assumptions, except for the knowledge of the giant component). This combined with Proposition~\ref{Prop_twocore_size_lower} will give the in-probability convergence of the relative size of the giant $2$-core, i.e., the following theorem.

\begin{theorem}\label{Thm_main_two_core_max}
For any $\eps > 0$ and $p\ne p_c$, $\frac{1}{n}|\Ctwomax(G_n(p))| \overset{p}{\to} \zetatwoinfty(p)$. At $p=p_c$, the right hand side serves as a high-probability upper bound.
\end{theorem}

An analogous result holds for the $2$-core of the entire graph $G_n(p)$, which will be detailed in Appendix~\ref{appendix_full_two_core}.  
In this appendix, we prove the following lemma.

\begin{lemma}\label{Lemma_twocore_size_upper}
For any $\eps > 0$ and $p\in(0,1]$, we have
\begin{equation}\label{Eqn_lemma_twocore_size_upper_main_twocore_max}
    \frac{1}{n}|\Ctwo(\Cmax(G_n(p)))| \le \zetatwoinfty(p) + \eps \qquad\text{ whp.}
\end{equation}
\end{lemma}
\begin{proof}
Note that this lemma becomes trivial when $\frac{1}{n}|\Cmax(G_n(p))| < \eps$. Thus, we may assume $|\Cmax(G_n(p))| \ge \eps n$.

Recall from Section~\ref{Sec_two_core_exists} the definition of the localized event that $o\in \Ctwoell(G)$ for a rooted graph $(G,o)$; in particular, we have $\Ctwoinfty(G) = \bigcap_{\ell\in\N} \Ctwoell(G)$. Let $\zetatwoell(p) = \Pp_{(G,o)}(o\in\Ctwoell(G(p)))$, which approaches $\zetatwoinfty(p) = \Pp_{(G,o)}(o\in\Ctwoinfty(G(p)))$ from above. By choosing $\ell$ sufficiently large, we can guarantee that $\zetatwoell(p) \le \zetatwoinfty(p) + \eps / 3$.

Since $\Ctwoell$ describes a local event, by Lemma~\ref{Lemma_3_1_local_event_func_in_prob}, the local convergence of $G_n$ implies
\begin{equation}
    \frac{1}{n}\sum_{v\in V(G_n)} \mathbbm{1}\{v \in \Ctwoell(G_n(p))\} = \frac{1}{n}|\Ctwoell(G_n(p))| \overset{p}{\to} \zetatwoell(p).
\end{equation}
In particular,
\begin{equation}\label{Eqn_lemma_twocore_size_upper_zetatwoell_upper}
    \frac{1}{n}|\Ctwoell(G_n(p))| \le \zetatwoell(p) + \frac{\eps}{3} \le \zetatwoinfty(p) + \frac{2\eps}{3} \qquad\text{ whp}.
\end{equation}

It now suffices to bound $\Ctwo(\Cmax(G_n(p))) \backslash \Ctwoell(G_n(p))$. Observe that a vertex in $\Ctwo(G_n(p))$ but not in $\Ctwoell(G_n(p))$ must have an $\ell$-neighborhood that contains a full connected component of $G_n(p)$. In particular, $v \in \Ctwo(\Cmax(G_n(p))) \backslash \Ctwoell(G_n(p))$ implies $|\cN_\ell(v; G_n)| \ge |\Cmax(G_n(p))| \ge \eps n$. By Lemma~\ref{Lemma_bounded_nbrh_size}, we conclude that
\begin{equation}
    \frac{1}{n} |\Ctwo(\Cmax(G_n(p))) \backslash \Ctwoell(G_n(p))| \le \frac{1}{n} \sum_{v\in V(G_n)} \mathbbm{1}\{|\cN_\ell(v)| \ge \eps n\} \le \frac{\eps}{3} \qquad\text{ whp}.
\end{equation}
Combining this with \eqref{Eqn_lemma_twocore_size_upper_zetatwoell_upper} concludes the proof.
\end{proof}

\section{Size of the entire $2$-core}\label{appendix_full_two_core}

We now examine the size of the entire $2$-core of $G_n(p)$,
proving the following theorem.

\begin{theorem}\label{Thm_main_two_core_entire}
For any $\eps > 0$ and $p\ne p_c$, $\frac{1}{n}|\Ctwo(G_n(p))| \overset{p}{\to} \zetatwo(p)$. At $p=p_c$, the right hand side serves as a high-probability upper bound.
\end{theorem}

Since we already know from Theorem~\ref{Thm_main_two_core_max} the in-probability convergence of the relative size $\Ctwo(\Cmax(G_n(p)))$ to $\zetatwoinfty(p)$ when $p \ne p_c$, we may simply focus on the quantity $\Ctwo(G_n(p) \backslash \Cmax(G_n(p)))$, i.e., the $2$-core outside the largest connected component of $G_n(p)$.

It suffices to show the following result, from which Theorem~\ref{Thm_main_two_core_entire} follows.

\begin{proposition}
    For any $p\in[0,1]$,
    \begin{equation}
        \frac{1}{n}|\Ctwo(G_n(p) \backslash \Cmax(G_n(p)))| \overset{p}{\to} \zetatwo(p) - \zetatwoinfty(p).
    \end{equation}
\end{proposition}
\begin{proof}
    Given a graph $G$, let $\zetatwolessinfty(p) = \zetatwo(p) - \zetatwoinfty(p)$, and let $Z_{<\ell}(G)$ (resp. $Z_{\ge\ell}(G)$) denote the set of vertices in connected components of $G$ with size less than $\ell$ (resp. no less than $\ell$).

    Fix some $\eps > 0$. By Corollary~\ref{Cor_bound_k_large_components_outside_giant}, there exists a sufficiently large constant $\ell$ such that
    \begin{equation}
        \frac{1}{n}|Z_{\ge \ell}(G_n(p)) \backslash \Cmax(G_n(p))| \le \frac{\eps}{3} \qquad\text{ whp}. 
    \end{equation}
    Observe that, when $|\Cmax(G_n(p))| \ge \ell$, we have
    \begin{equation*}
        G_n(p) \backslash \Cmax(G_n(p)) = Z_{< \ell}(G_n(p)) \sqcup \big(Z_{\ge \ell}(G_n(p)) \backslash \Cmax(G_n(p))\big)
    \end{equation*}
    with $\sqcup$ denoting disjoint union,
    and when $|\Cmax(G_n(p))| < \ell$, $\Ctwo(G_n(p)\backslash \Cmax(G_n(p)))$ differs from $\Ctwo(Z_{< \ell}(G_n(p)))$ by at most $\ell$ vertices. In either case, it suffices to show that, by choosing $\ell$ sufficiently large, we can guarantee
    \begin{equation}\label{Eqn_proof_prop_entire_two_core_reduces_to}
        \left| \frac{1}{n} |\Ctwo(Z_{< \ell}(G_n(p)))| - \zetatwolessinfty(p) \right| \le \frac{2\eps}{3} \qquad\text{ whp}.
    \end{equation}

    Define $\zetatwolessell(p):= \Pp_{(G,o)\sim\mu}(o\in\Ctwo(G(p)), |C(o;G(p))| < \ell)$. Note that the corresponding events form a monotone increasing sequence as $\ell\to\infty$, and hence, with $(G,o)\sim\mu$,
    \begin{equation*}
        \lim_{\ell\to\infty}\zetatwolessell(p) = \lim_{\ell\to\infty}\Pp(o\in\Ctwo(G(p)), |C(o;G(p))| < \ell) = \Pp(o\in\Ctwo(G(p)), |C(o;G(p))| < \infty) = \zetatwolessinfty(p).
    \end{equation*}
    By choosing $\ell$ sufficiently large, we can guarantee $|\zetatwolessell(p) - \zetatwolessinfty(p)| \le \eps / 3$. Further, the event that $o\in\Ctwo(G)$ and $|C(o;G)| < \ell$ for a rooted graph $(G,o)$ is local, and by Lemma~\ref{Lemma_3_1_local_event_func_in_prob},
    \begin{equation}
        \frac{1}{n}|\Ctwo(Z_{< \ell}(G_n(p)))| = \frac{1}{n} \sum_{v\in V(G_n)} \mathbbm{1}\{v\in\Ctwo(G_n(p)), |C(v;G_n(p))|<\ell\} \overset{p}{\to} \zetatwolessell(p).
    \end{equation}
    Thus, \eqref{Eqn_proof_prop_entire_two_core_reduces_to} holds and our proof is complete.
\end{proof}

\section{Proof details for Section~\ref{Sec_two_core_size_conv}}\label{Appendix_proofs}

\subsection{Proof details for Section~\ref{Sec_two_core_exists}}
\label{app:infinite-2core}

We first establish a general lemma that lower bounds the probability that a local event for the root using that for any vertex in a neighborhood of the root, and essentially suggests that the previous lemma holds with any choice of $r\in\N$. This is a direct consequence of the local convergence (in probability) and the submodularity of the limit.

\begin{lemma}\label{Lemma_local_event_probability_upper_bounded_by_nbrh_event_probability}
    Let $\cE(G,o)$ be a local event. Then for any $\delta > 0$, $r\in\N$, and $p\in[0,1]$, there exists a constant $D \in \N$ (depending on $\delta$ and $r$) such that
    \begin{equation}\label{Eqn_lemma_local_event_prob_upper_bounded_by_nbhr_prob_main}
        \Pp(\exists u\in\cN_r(o; G) \text{ s.t. }\cE(G(p),u)) \le D \Pp(\cE(G(p),o)) + \delta.
    \end{equation}
    where all probabilities are understood as over the randomness of $(G,o)\sim\mu$ and the percolation.
\end{lemma}
\begin{proof}
    Observe that it suffices to show \eqref{Eqn_lemma_local_event_prob_upper_bounded_by_nbhr_prob_main} for $r=1$; for the general case, we simply use induction based on the fact that $\Pp(\exists u\in\cN_r(o;G) \text{ s.t. }\cE(G(p),u)) = \Pp(\exists v \in\cN_1(o;G), u\in\cN_{r-1}(v;G) \text{ s.t. }\cE(G(p),u))$.
    
    Since $\mu$ is a distribution on the space of locally finite graphs, we have
    \begin{equation*}
        \deg(o;G)\, \vee \,\max_{u\sim_G o}\; \deg(u;G) < \infty \qquad\text{ a.s.},
    \end{equation*}
    where the notation $u\sim v$ indicates the adjacency of two vertices $u,v\in V$ (in $G$).
    Thus, for any $\delta > 0$, there exists $D < \infty$ such that $\deg(o) < D$ and $\max_{u\sim o} \deg(u;G) < D$ with probability at least $1-\delta$.

    Further, $\mu$ as a local limit is unimodular and hence involution invariant \cite[Proposition~2.2]{aldous2007processes}. In particular, this implies that
    \begin{equation}
        \E\left[\sum_{u\sim o} \Pp(\cE(G(p),o), \;\deg(o)<D|(G,o))\right] = \E\left[\sum_{u\sim o} \Pp(\cE(G(p),u), \;\deg(u)<D|(G,o))\right].
    \end{equation}
    (Note that this depends on the fact that $\Pp(\cE(G(p),o), \;\deg(o)<D|(G,o))$ is a local function.)
    The right hand side is an upper bound for $\Pp(\exists u\sim o \text{ s.t. } \cE(G,u), \;\deg(u)<D)$. Thus,
    \begin{align*}
        \Pp\big(\exists u\sim o \text{ s.t. } \cE(G(p),u)\big) &\le \Pp(\exists u\sim o \text{ s.t. } \deg(u) \ge D) + \E\left[\deg(o) \cdot \Pp(\cE(G(p),o), \;\deg(o)<D|(G,o))\right] \\
        &\le \delta + D \Pp(\cE(G(p),o))
    \end{align*}
    as we claimed.
\end{proof}

Lemma~\ref{Lemma_local_event_probability_upper_bounded_by_nbrh_event_probability} essentially allows us to bound the probability that $o$ lies in an infinite component of $\Ctwo(G(p))$ by the probability that it lies in a bounded neighborhood of an infinite component of $\Ctwo(G(p))$.
To state this formally, we define the $r$-neighborhood 
 $\cN_r(A;G)$ of a set of vertices $A$ as the induced subgraph on the set of vertices with distance at most $r$ from $A$.

\begin{corollary}\label{Cor_relate_Pr_in_infty_two_core_to_Pr_in_finite_nbrh_of_twocore} For $p>p_c$ and any $r\in\N$, the probability  
$\Pp(o\in \Ctwoinfty(G(p))) $ is non-zero if and only if
$ \Pp(o \in \cN_r(\Ctwoinfty(G(p))))$ is non-zero.
\end{corollary}

\begin{proof}
    The ``only if'' direction is trivial. 
    For the ``if'' direction, fix $p > p_c$ and assume $a := \Pp(o \in \cN_r(\Ctwoinfty(G(p)))) > 0$. Since $\Ctwoinfty(G(p)) \subseteq \Ctwoell(G(p))$ for any $\ell\in\N$ a.s., we have
    \begin{equation}
        \Pp(o \in \cN_r(\Ctwoell(G(p)))) \ge \Pp(o \in \cN_r(\Ctwoinfty(G(p)))) = a.
    \end{equation}
    By Lemma~\ref{Lemma_local_event_probability_upper_bounded_by_nbrh_event_probability}, there exists $D\in\N$ such that
    \begin{equation}
        \Pp(o \in \Ctwoell(G(p))) \ge \frac{1}{D} \Big( \Pp(o \in \cN_r(\Ctwoell(G(p)))) - \frac{a}{2} \Big) \ge \frac{a}{2D}.
    \end{equation}
    Taking the limit as $\ell\to\infty$ gives $\zetatwoinfty(p) = \Pp(o \in \Ctwoinfty(G(p))) = \lim_{\ell\to\infty} \Pp(o \in \Ctwoell(G(p))) \ge \frac{a}{2D} > 0$.
\end{proof}
We now show that, for $(G,o)\sim\mu$, the probability of $o$ lying in some finite neighborhood of $\Ctwoinfty(G(p))$ is strictly positive.

\begin{lemma}\label{Lemma_positive_probability_in_finite_nbrh_of_2_core}
    For any $p > p_c$, there exist some $r\in\N$ such that $\Pp\big(o\in\cN_r(\Ctwoinfty(G(p)))\big) > 0$. %
\end{lemma}
\begin{proof}
    It suffices to show that
    \begin{equation}\label{Eqn_proof_lemma_positive_prob_in_nbrh_of_two_core_suffice_to_show}
        \Pp\bigg(o\in\bigcup_{r\in\N}\cN_r(\Ctwoinfty(G(p)))\bigg) = \Pp(|C(o;G(p))|=\infty),
    \end{equation}
    which is positive due to the super-criticality assumption. The lemma follows from the fact that the quantity above is the monotone limit of $\Pp(o\in\cN_r(\Ctwoinfty(G(p))))$ as $r\to \infty$.

    If $|C(o;G(p))|=\infty$ yet $o\notin \cN_r\big(\Ctwoinfty(G(p))\big)$ for all $r\in\N$, then the $2$-core of this infinite component must be empty, and thus there exists a unique ray extending from $o$ to infinity. Let $\mathcal{E}_p$ denote such an event. \eqref{Eqn_proof_lemma_positive_prob_in_nbrh_of_two_core_suffice_to_show} reduces to the claim that $\Pp(\mathcal{E}_p) = 0$.
    
    Assume that $\Pp(\mathcal{E}_p) = \eps > 0$.
    Choose $p' \in(p_c,p)$ such that $\zeta(p') > \zeta(p) - \eps$, which is possible due to continuity of $\zeta$ above $p_c$ (see Proposition~\ref{Prop_cont_zeta}). Obtain $G(p')$ by keeping the common edges in $G(p)$ and $G(p'/p)$; equivalently, first obtain $G(p)$ and then do percolation with probability $p'/p$ on it.
    Conditional on $\mathcal{E}_p$, there is a unique ray in $G(p)$ from $o$ to infinity, along which the removal of any edge leaves $o$ in a finite component, and thus, with probability $1$, $o$ is in a finite component in $G(p')$. Thus, $\zeta(p') \le \zeta(p) - \eps$, a contradiction.
\end{proof}

The proof of Proposition~\ref{Prop_emerge_two_core} is now a $3$-line argument.

\begin{proof}[Proof of Proposition~\ref{Prop_emerge_two_core}]
    The second part is trivial, since $\zeta(p)=0$ for $p<p_c$. Assume thus that $p>p_c$.  By Lemma~\ref{Lemma_positive_probability_in_finite_nbrh_of_2_core},  there exist some $r\in\N$ such that $\Pp\big(o\in\cN_r(\Ctwoinfty(G(p)))\big) > 0$, which by Corollary~\ref{Cor_relate_Pr_in_infty_two_core_to_Pr_in_finite_nbrh_of_twocore} implies that
$\zetatwoinfty(p)=\Pp(o\in \Ctwoinfty(G(p)))>0 $.
\end{proof}

Finally, we provide the omitted proof for Lemma~\ref{Lemma_Ztwoinfty_k_of_C1_at_p_close_to_limit}. 
We start with
the following lemma,
which follows from the local convergence of $G_n(p)$ to percolation in the limit (see Lemma~\ref{Lemma_3_1_local_event_func_in_prob}) and the fact that $\Ctwoell$ can be determined locally. 

    \begin{lemma}\label{Lemma_inp_conv_of_Z_two_ell}
        For any $p\in[0,1]$ and any $\ell\in\N$, $\frac{1}{n} |\Ctwoell(G_n(p))| \overset{p}{\to} \zetatwoell(p):= \Pp_{(G,o)\sim\mu}(o \in \Ctwoell(G(p)))$.
    \end{lemma}

Next we prove Lemma~\ref{Lemma_Ztwoinfty_k_of_C1_at_p_close_to_limit}, which was used in Section~\ref{Sec_two_core_exists} to prove Proposition~\ref{Cor_exist_giant_two_core}.

\begin{proof}[Proof of Lemma~\ref{Lemma_Ztwoinfty_k_of_C1_at_p_close_to_limit}]

    Observe that $\Ctwoell(G(p)) \supseteq \Ctwoinfty(G(p))$. Thus, $\zetatwoell(p) \ge \zetatwoinfty(p)$, which is positive for $p>p_c$ due to Proposition~\ref{Prop_emerge_two_core}. By Lemma~\ref{Lemma_inp_conv_of_Z_two_ell}, we have $\frac{1}{n} |\Ctwoell(G_n(p))| \overset{p}{\to} \zetatwoell(p)$. It suffices to show that, for $\ell$ sufficiently large,
    \begin{equation}
        \frac{1}{n} |\Ctwoell(\Cmax(G_n(p)))| \ge \frac{1}{n} |\Ctwoell(G_n(p))| - \eps \qquad\text{ whp},
    \end{equation}
    or equivalently,
    \begin{equation}\label{Eqn_proof_Lemma_Ztwoinfty_k_of_C1_at_p_close_to_limit_STP}
        \frac{1}{n} \Big|\Ctwoell\big(G_n(p)\backslash \Cmax(G_n(p))\big)\Big| \le  \eps \qquad\text{ whp},
    \end{equation}
    Note that any vertex $v$ in $\Ctwoell\big(G_n(p)\backslash \Cmax(G_n(p))\big)$ must be in a connected component of size at least $\ell$ but not the largest component. By Corollary~\ref{Cor_bound_k_large_components_outside_giant}, \eqref{Eqn_proof_Lemma_Ztwoinfty_k_of_C1_at_p_close_to_limit_STP} holds for $\ell$ sufficiently large.
\end{proof}

\newpage

\subsection{Illustrations for Section~\ref{Sec_two_core_size_lower}}
\label{app:extra figures}
\nobreak
\begin{figure}[H]
  \centering

  \begin{subfigure}[b]{0.9\textwidth}
  \centering
  
\begin{tikzpicture}[scale=0.4]
\coordinate (H1) at (0,0);
\coordinate (H2) at (0,2);

\coordinate (S17) at ({sqrt(3)},-1);
\coordinate (S16) at ({2*sqrt(3)},0);
\coordinate (S15) at ({2*sqrt(3)},2);
\coordinate (S14) at ({3*sqrt(3)},3);
\coordinate (S13) at ({4*sqrt(3)},2);
\coordinate (S12) at ({4*sqrt(3)},0);
\coordinate (S11) at ({5*sqrt(3)},-1);
\coordinate (S10) at ({6*sqrt(3)},0);
\coordinate (S9) at ({6*sqrt(3)},2);
\coordinate (S8) at ({7*sqrt(3)},3);
\coordinate (S7) at ({8*sqrt(3)},2);
\coordinate (S6) at ({9*sqrt(3)},3);
\coordinate (S5) at ({10*sqrt(3)},2);
\coordinate (S4) at ({11*sqrt(3)},3);
\coordinate (S3) at ({12*sqrt(3)},2);
\coordinate (S2) at ({13*sqrt(3)},3);
\coordinate (S1) at ({14*sqrt(3)},2);

\coordinate (Br1N3) at ({11*sqrt(3)},5);
\coordinate (Br1N2) at ({12*sqrt(3)},6);
\coordinate (Br1N1) at ({13*sqrt(3)},5);

\coordinate (Br2N9) at ({3*sqrt(3)},-1);
\coordinate (Br2N8) at ({3*sqrt(3)},-3);
\coordinate (Br2N7) at ({4*sqrt(3)},-4);
\coordinate (Br2N6) at ({5*sqrt(3)},-3);
\coordinate (Br2N5) at ({6*sqrt(3)},-4);
\coordinate (Br2N4) at ({7*sqrt(3)},-3);
\coordinate (Br2N3) at ({8*sqrt(3)},-4);
\coordinate (Br2N2) at ({9*sqrt(3)},-3);
\coordinate (Br2N1) at ({10*sqrt(3)},-4);

\coordinate (Br3N3) at ({10*sqrt(3)},0);
\coordinate (Br3N2) at ({11*sqrt(3)},-1);
\coordinate (Br3N1) at ({12*sqrt(3)},0);
\coordinate (Br3N1p) at ({11*sqrt(3)},-3);

\coordinate (Br4N2) at ({8*sqrt(3)},0);
\coordinate (Br4N1p1) at ({9*sqrt(3)},1);
\coordinate (Br4N1p2) at ({9*sqrt(3)},-1);
\coordinate (Br4N1p3) at ({8*sqrt(3)},-2);
\coordinate (Br4N1p4) at ({7*sqrt(3)},-1);

\coordinate (Br5N2) at ({sqrt(3)},3);
\coordinate (Br5N1p1) at ({sqrt(3)},5);
\coordinate (Br5N1p2) at ({2*sqrt(3)},4);

\draw[->, >=latex] (S16) -- (S17);
\draw[->, >=latex] (S15) -- (S16);
\draw[->, >=latex] (S14) -- (S15);
\draw[->, >=latex] (S13) -- (S14);
\draw[->, >=latex] (S12) -- (S13);
\draw[->, >=latex] (S11) -- (S12);
\draw[->, >=latex] (S10) -- (S11);
\draw[->, >=latex] (S9) -- (S10);
\draw[->, >=latex] (S8) -- (S9);
\draw[->, >=latex] (S7) -- (S8);
\draw[->, >=latex] (S6) -- (S7);
\draw[->, >=latex] (S5) -- (S6);
\draw[->, >=latex] (S4) -- (S5);
\draw[->, >=latex] (S3) -- (S4);
\draw[->, >=latex] (S2) -- (S3);
\draw[->, >=latex] (S1) -- (S2);

\draw[->, >=latex] (Br1N1) -- (Br1N2);
\draw[->, >=latex] (Br1N2) -- (Br1N3);
\draw[->, >=latex] (Br1N3) -- (S4);

\draw[->, >=latex] (Br2N1) -- (Br2N2);
\draw[->, >=latex] (Br2N2) -- (Br2N3);
\draw[->, >=latex] (Br2N3) -- (Br2N4);
\draw[->, >=latex] (Br2N4) -- (Br2N5);
\draw[->, >=latex] (Br2N5) -- (Br2N6);
\draw[->, >=latex] (Br2N6) -- (Br2N7);
\draw[->, >=latex] (Br2N7) -- (Br2N8);
\draw[->, >=latex] (Br2N8) -- (Br2N9);
\draw[->, >=latex] (Br2N9) -- (S12);

\draw[->, >=latex] (Br3N1) -- (Br3N2);
\draw[->, >=latex] (Br3N1p) -- (Br3N2);
\draw[->, >=latex] (Br3N2) -- (Br3N3);
\draw[->, >=latex] (Br3N3) -- (S5);

\draw[->, >=latex] (Br4N1p1) -- (Br4N2);
\draw[->, >=latex] (Br4N1p2) -- (Br4N2);
\draw[->, >=latex] (Br4N1p3) -- (Br4N2);
\draw[->, >=latex] (Br4N1p4) -- (Br4N2);
\draw[->, >=latex] (Br4N2) -- (S7);

\draw[->, >=latex] (Br5N1p1) -- (Br5N2);
\draw[->, >=latex] (Br5N1p2) -- (Br5N2);

\draw[dashed] (Br5N2) -- (H2);
\draw[dashed] (S17) -- (H1);

\node[circle, draw=purple, fill=purple, inner sep=1pt, label=17] at (S17) {};
\node[circle, draw, fill=black, inner sep=1pt, label=below:16] at (S16) {};
\node[circle, draw, fill=black, inner sep=1pt, label=15] at (S15) {};
\node[circle, draw, fill=black, inner sep=1pt, label=14] at (S14) {};
\node[circle, draw, fill=black, inner sep=1pt, label=13] at (S13) {};
\node[circle, draw=purple, fill=purple, inner sep=1pt, label=above left:12] at (S12) {};
\node[circle, draw=purple, fill=purple, inner sep=1pt, label=11] at (S11) {};
\node[circle, draw=purple, fill=purple, inner sep=1pt, label=above left:10] at (S10) {};
\node[circle, draw=purple, fill=purple, inner sep=1pt, label=9] at (S9) {};
\node[circle, draw, fill=black, inner sep=1pt, label=8] at (S8) {};
\node[circle, draw, fill=black, inner sep=1pt, label=7] at (S7) {};
\node[circle, draw, fill=black, inner sep=1pt, label=6] at (S6) {};
\node[circle, draw, fill=black, inner sep=1pt, label=5] at (S5) {};
\node[circle, draw=red, fill=red, inner sep=1pt, label={[text=red]below:4}] at (S4) {};
\node[circle, draw=red, fill=red, inner sep=1pt, label={[text=red]3}] at (S3) {};
\node[circle, draw=red, fill=red, inner sep=1pt, label={[text=red]below:2}] at (S2) {};
\node[circle, draw=red, fill=red, inner sep=1pt, label={[text=red]below:1}] at (S1) {};

\node[circle, draw=red, fill=red, inner sep=1pt, label={[text=red]3}] at (Br1N3) {};
\node[circle, draw=red, fill=red, inner sep=1pt, label={[text=red]2}] at (Br1N2) {};
\node[circle, draw=red, fill=red, inner sep=1pt, label={[text=red]1}] at (Br1N1) {};

\node[circle, draw=red, fill=red, inner sep=1pt, label={[text=red]below:1}] at (Br2N1) {};
\node[circle, draw=red, fill=red, inner sep=1pt, label={[text=red]below:2}] at (Br2N2) {};
\node[circle, draw=red, fill=red, inner sep=1pt, label={[text=red]below:3}] at (Br2N3) {};
\node[circle, draw=red, fill=red, inner sep=1pt, label={[text=red]below:4}] at (Br2N4) {};
\node[circle, draw, fill=black, inner sep=1pt, label=below:5] at (Br2N5) {};
\node[circle, draw, fill=black, inner sep=1pt, label=below:6] at (Br2N6) {};
\node[circle, draw, fill=black, inner sep=1pt, label=below:7] at (Br2N7) {};
\node[circle, draw, fill=black, inner sep=1pt, label=below:8] at (Br2N8) {};
\node[circle, draw=purple, fill=purple, inner sep=1pt, label=9] at (Br2N9) {};

\node[circle, draw=red, fill=red, inner sep=1pt, label={[text=red]below:3}] at (Br3N3) {};
\node[circle, draw=red, fill=red, inner sep=1pt, label={[text=red]2}] at (Br3N2) {};
\node[circle, draw=red, fill=red, inner sep=1pt, label={[text=red]below:1}] at (Br3N1) {};
\node[circle, draw=red, fill=red, inner sep=1pt, label={[text=red]below:1}] at (Br3N1p) {};

\node[circle, draw=red, fill=red, inner sep=1pt, label={[text=red]above left:2}] at (Br4N2) {};
\node[circle, draw=red, fill=red, inner sep=1pt, label={[text=red]1}] at (Br4N1p1) {};
\node[circle, draw=red, fill=red, inner sep=1pt, label={[text=red]below:1}] at (Br4N1p2) {};
\node[circle, draw=red, fill=red, inner sep=1pt, label={[text=red]left:1}] at (Br4N1p3) {};
\node[circle, draw=red, fill=red, inner sep=1pt, label={[text=red]left:1}] at (Br4N1p4) {};

\node[circle, draw=red, fill=red, inner sep=1pt, label={[text=red]below:2}] at (Br5N2) {};
\node[circle, draw=red, fill=red, inner sep=1pt, label={[text=red]1}] at (Br5N1p1) {};
\node[circle, draw=red, fill=red, inner sep=1pt, label={[text=red]1}] at (Br5N1p2) {};

\draw[dashed] (-1,{1-sqrt(3)}) arc (-60:60:2cm) node[above] {$H$};
\end{tikzpicture}

\caption{Color vertices purple or red according to their depth modulo $2\ell$.\,\,\phantom{xxxxxxxxxxxxxx}\phantom{xxxxxxxxxx}}
\end{subfigure}

\begin{subfigure}[b]{0.9\textwidth}
  \centering
  
\begin{tikzpicture}[scale=0.40]
\coordinate (H1) at (0,0);
\coordinate (H2) at (0,2);

\coordinate (S17) at ({sqrt(3)},-1);
\coordinate (S16) at ({2*sqrt(3)},0);
\coordinate (S15) at ({2*sqrt(3)},2);
\coordinate (S14) at ({3*sqrt(3)},3);
\coordinate (S13) at ({4*sqrt(3)},2);
\coordinate (S12) at ({4*sqrt(3)},0);
\coordinate (S11) at ({5*sqrt(3)},-1);
\coordinate (S10) at ({6*sqrt(3)},0);
\coordinate (S9) at ({6*sqrt(3)},2);
\coordinate (S8) at ({7*sqrt(3)},3);
\coordinate (S7) at ({8*sqrt(3)},2);
\coordinate (S6) at ({9*sqrt(3)},3);
\coordinate (S5) at ({10*sqrt(3)},2);
\coordinate (S4) at ({11*sqrt(3)},3);
\coordinate (S3) at ({12*sqrt(3)},2);
\coordinate (S2) at ({13*sqrt(3)},3);
\coordinate (S1) at ({14*sqrt(3)},2);

\coordinate (Br1N3) at ({11*sqrt(3)},5);
\coordinate (Br1N2) at ({12*sqrt(3)},6);
\coordinate (Br1N1) at ({13*sqrt(3)},5);

\coordinate (Br2N9) at ({3*sqrt(3)},-1);
\coordinate (Br2N8) at ({3*sqrt(3)},-3);
\coordinate (Br2N7) at ({4*sqrt(3)},-4);
\coordinate (Br2N6) at ({5*sqrt(3)},-3);
\coordinate (Br2N5) at ({6*sqrt(3)},-4);
\coordinate (Br2N4) at ({7*sqrt(3)},-3);
\coordinate (Br2N3) at ({8*sqrt(3)},-4);
\coordinate (Br2N2) at ({9*sqrt(3)},-3);
\coordinate (Br2N1) at ({10*sqrt(3)},-4);

\coordinate (Br3N3) at ({10*sqrt(3)},0);
\coordinate (Br3N2) at ({11*sqrt(3)},-1);
\coordinate (Br3N1) at ({12*sqrt(3)},0);
\coordinate (Br3N1p) at ({11*sqrt(3)},-3);

\coordinate (Br4N2) at ({8*sqrt(3)},0);
\coordinate (Br4N1p1) at ({9*sqrt(3)},1);
\coordinate (Br4N1p2) at ({9*sqrt(3)},-1);
\coordinate (Br4N1p3) at ({8*sqrt(3)},-2);
\coordinate (Br4N1p4) at ({7*sqrt(3)},-1);

\coordinate (Br5N2) at ({sqrt(3)},3);
\coordinate (Br5N1p1) at ({sqrt(3)},5);
\coordinate (Br5N1p2) at ({2*sqrt(3)},4);

\draw[->, >=latex] (S16) -- (S17);
\draw[->, >=latex] (S15) -- (S16);
\draw[->, >=latex] (S14) -- (S15);
\draw[->, >=latex] (S13) -- (S14);
\draw[->, >=latex] (S12) -- (S13);
\draw[->, >=latex, purple] (S11) -- (S12);
\draw[->, >=latex, purple] (S10) -- (S11);
\draw[->, >=latex, purple] (S9) -- (S10);
\draw[->, >=latex] (S8) -- (S9);
\draw[->, >=latex] (S7) -- (S8);
\draw[->, >=latex] (S6) -- (S7);
\draw[->, >=latex] (S5) -- (S6);
\draw[->, >=latex] (S4) -- (S5);
\draw[->, >=latex, red] (S3) -- (S4);
\draw[->, >=latex, red] (S2) -- (S3);
\draw[->, >=latex, red] (S1) -- (S2);

\draw[->, >=latex, red] (Br1N1) -- (Br1N2);
\draw[->, >=latex, red] (Br1N2) -- (Br1N3);
\draw[->, >=latex, red] (Br1N3) -- (S4);

\draw[->, >=latex, red] (Br2N1) -- (Br2N2);
\draw[->, >=latex, red] (Br2N2) -- (Br2N3);
\draw[->, >=latex, red] (Br2N3) -- (Br2N4);
\draw[->, >=latex] (Br2N4) -- (Br2N5);
\draw[->, >=latex] (Br2N5) -- (Br2N6);
\draw[->, >=latex] (Br2N6) -- (Br2N7);
\draw[->, >=latex] (Br2N7) -- (Br2N8);
\draw[->, >=latex] (Br2N8) -- (Br2N9);
\draw[->, >=latex] (Br2N9) -- (S12);

\draw[->, >=latex, red] (Br3N1) -- (Br3N2);
\draw[->, >=latex, red] (Br3N1p) -- (Br3N2);
\draw[->, >=latex, red] (Br3N2) -- (Br3N3);
\draw[->, >=latex] (Br3N3) -- (S5);

\draw[->, >=latex, red] (Br4N1p1) -- (Br4N2);
\draw[->, >=latex, red] (Br4N1p2) -- (Br4N2);
\draw[->, >=latex, red] (Br4N1p3) -- (Br4N2);
\draw[->, >=latex, red] (Br4N1p4) -- (Br4N2);
\draw[->, >=latex] (Br4N2) -- (S7);

\draw[->, >=latex, red] (Br5N1p1) -- (Br5N2);
\draw[->, >=latex, red] (Br5N1p2) -- (Br5N2);

\draw[dashed] (Br5N2) -- (H2);
\draw[dashed] (S17) -- (H1);

\node[circle, draw=purple, fill=purple, inner sep=1pt, label={[text=purple]17}] at (S17) {};
\node[circle, draw, fill=black, inner sep=1pt, label=below:16] at (S16) {};
\node[circle, draw, fill=black, inner sep=1pt, label=15] at (S15) {};
\node[circle, draw, fill=black, inner sep=1pt, label=14] at (S14) {};
\node[circle, draw, fill=black, inner sep=1pt, label=13] at (S13) {};
\node[circle, draw=purple, fill=purple, inner sep=1pt, label={[text=purple]above left:12}] at (S12) {};
\node[circle, draw=purple, fill=purple, inner sep=1pt, label={[text=purple]11}] at (S11) {};
\node[circle, draw=purple, fill=purple, inner sep=1pt, label={[text=purple]above left:10}] at (S10) {};
\node[circle, draw=purple, fill=purple, inner sep=1pt, label={[text=purple]9}] at (S9) {};
\node[circle, draw, fill=black, inner sep=1pt, label=8] at (S8) {};
\node[circle, draw, fill=black, inner sep=1pt, label=7] at (S7) {};
\node[circle, draw, fill=black, inner sep=1pt, label=6] at (S6) {};
\node[circle, draw, fill=black, inner sep=1pt, label=5] at (S5) {};
\node[circle, draw=red, fill=red, inner sep=1pt, label={[text=red]below:4}] at (S4) {};
\node[circle, draw=red, fill=red, inner sep=1pt, label={[text=red]3}] at (S3) {};
\node[circle, draw=red, fill=red, inner sep=1pt, label={[text=red]below:2}] at (S2) {};
\node[circle, draw=red, fill=red, inner sep=1pt, label={[text=red]below:1}] at (S1) {};

\node[circle, draw=red, fill=red, inner sep=1pt, label={[text=red]3}] at (Br1N3) {};
\node[circle, draw=red, fill=red, inner sep=1pt, label={[text=red]2}] at (Br1N2) {};
\node[circle, draw=red, fill=red, inner sep=1pt, label={[text=red]1}] at (Br1N1) {};

\node[circle, draw=red, fill=red, inner sep=1pt, label={[text=red]below:1}] at (Br2N1) {};
\node[circle, draw=red, fill=red, inner sep=1pt, label={[text=red]below:2}] at (Br2N2) {};
\node[circle, draw=red, fill=red, inner sep=1pt, label={[text=red]below:3}] at (Br2N3) {};
\node[circle, draw=red, fill=red, inner sep=1pt, label={[text=red]below:4}] at (Br2N4) {};
\node[circle, draw, fill=black, inner sep=1pt, label=below:5] at (Br2N5) {};
\node[circle, draw, fill=black, inner sep=1pt, label=below:6] at (Br2N6) {};
\node[circle, draw, fill=black, inner sep=1pt, label=below:7] at (Br2N7) {};
\node[circle, draw, fill=black, inner sep=1pt, label=below:8] at (Br2N8) {};
\node[circle, draw=purple, fill=purple, inner sep=1pt, label={[text=purple]9}] at (Br2N9) {};

\node[circle, draw=red, fill=red, inner sep=1pt, label={[text=red]below:3}] at (Br3N3) {};
\node[circle, draw=red, fill=red, inner sep=1pt, label={[text=red]2}] at (Br3N2) {};
\node[circle, draw=red, fill=red, inner sep=1pt, label={[text=red]below:1}] at (Br3N1) {};
\node[circle, draw=red, fill=red, inner sep=1pt, label={[text=red]below:1}] at (Br3N1p) {};

\node[circle, draw=red, fill=red, inner sep=1pt, label={[text=red]above left:2}] at (Br4N2) {};
\node[circle, draw=red, fill=red, inner sep=1pt, label={[text=red]1}] at (Br4N1p1) {};
\node[circle, draw=red, fill=red, inner sep=1pt, label={[text=red]below:1}] at (Br4N1p2) {};
\node[circle, draw=red, fill=red, inner sep=1pt, label={[text=red]left:1}] at (Br4N1p3) {};
\node[circle, draw=red, fill=red, inner sep=1pt, label={[text=red]left:1}] at (Br4N1p4) {};

\node[circle, draw=red, fill=red, inner sep=1pt, label={[text=red]below:2}] at (Br5N2) {};
\node[circle, draw=red, fill=red, inner sep=1pt, label={[text=red]1}] at (Br5N1p1) {};
\node[circle, draw=red, fill=red, inner sep=1pt, label={[text=red]1}] at (Br5N1p2) {};

\draw[dashed] (-1,{1-sqrt(3)}) arc (-60:60:2cm) node[above] {$H$};
\end{tikzpicture}

\caption{Color edges between same-color vertices of consecutive depths to create segments.\phantom{xxxxxxxxxx}}
\end{subfigure}

\begin{subfigure}[b]{0.9\textwidth}
  \centering
  
\begin{tikzpicture}[scale=0.40]
\coordinate (H1) at (0,0);
\coordinate (H2) at (0,2);

\coordinate (S17) at ({sqrt(3)},-1);
\coordinate (S16) at ({2*sqrt(3)},0);
\coordinate (S15) at ({2*sqrt(3)},2);
\coordinate (S14) at ({3*sqrt(3)},3);
\coordinate (S13) at ({4*sqrt(3)},2);
\coordinate (S12) at ({4*sqrt(3)},0);
\coordinate (S11) at ({5*sqrt(3)},-1);
\coordinate (S10) at ({6*sqrt(3)},0);
\coordinate (S9) at ({6*sqrt(3)},2);
\coordinate (S8) at ({7*sqrt(3)},3);
\coordinate (S7) at ({8*sqrt(3)},2);
\coordinate (S6) at ({9*sqrt(3)},3);
\coordinate (S5) at ({10*sqrt(3)},2);
\coordinate (S4) at ({11*sqrt(3)},3);
\coordinate (S3) at ({12*sqrt(3)},2);
\coordinate (S2) at ({13*sqrt(3)},3);
\coordinate (S1) at ({14*sqrt(3)},2);

\coordinate (Br1N3) at ({11*sqrt(3)},5);
\coordinate (Br1N2) at ({12*sqrt(3)},6);
\coordinate (Br1N1) at ({13*sqrt(3)},5);

\coordinate (Br2N9) at ({3*sqrt(3)},-1);
\coordinate (Br2N8) at ({3*sqrt(3)},-3);
\coordinate (Br2N7) at ({4*sqrt(3)},-4);
\coordinate (Br2N6) at ({5*sqrt(3)},-3);
\coordinate (Br2N5) at ({6*sqrt(3)},-4);
\coordinate (Br2N4) at ({7*sqrt(3)},-3);
\coordinate (Br2N3) at ({8*sqrt(3)},-4);
\coordinate (Br2N2) at ({9*sqrt(3)},-3);
\coordinate (Br2N1) at ({10*sqrt(3)},-4);

\coordinate (Br3N3) at ({10*sqrt(3)},0);
\coordinate (Br3N2) at ({11*sqrt(3)},-1);
\coordinate (Br3N1) at ({12*sqrt(3)},0);
\coordinate (Br3N1p) at ({11*sqrt(3)},-3);

\coordinate (Br4N2) at ({8*sqrt(3)},0);
\coordinate (Br4N1p1) at ({9*sqrt(3)},1);
\coordinate (Br4N1p2) at ({9*sqrt(3)},-1);
\coordinate (Br4N1p3) at ({8*sqrt(3)},-2);
\coordinate (Br4N1p4) at ({7*sqrt(3)},-1);

\coordinate (Br5N2) at ({sqrt(3)},3);
\coordinate (Br5N1p1) at ({sqrt(3)},5);
\coordinate (Br5N1p2) at ({2*sqrt(3)},4);

\draw[->, >=latex] (S16) -- (S17);
\draw[->, >=latex] (S15) -- (S16);
\draw[->, >=latex] (S14) -- (S15);
\draw[->, >=latex] (S13) -- (S14);
\draw[->, >=latex] (S12) -- (S13);
\draw[->, >=latex, purple] (S11) -- (S12);
\draw[->, >=latex, purple] (S10) -- (S11);
\draw[->, >=latex, purple] (S9) -- (S10);
\draw[->, >=latex] (S8) -- (S9);
\draw[->, >=latex] (S7) -- (S8);
\draw[->, >=latex] (S6) -- (S7);
\draw[->, >=latex] (S5) -- (S6);
\draw[->, >=latex] (S4) -- (S5);
\draw[->, >=latex, red] (S3) -- (S4);
\draw[->, >=latex, red] (S2) -- (S3);
\draw[->, >=latex, red] (S1) -- (S2);

\draw[->, >=latex, red] (Br1N1) -- (Br1N2);
\draw[->, >=latex, red] (Br1N2) -- (Br1N3);
\draw[->, >=latex, red] (Br1N3) -- (S4);

\draw[->, >=latex, red] (Br2N1) -- (Br2N2);
\draw[->, >=latex, red] (Br2N2) -- (Br2N3);
\draw[->, >=latex, red] (Br2N3) -- (Br2N4);
\draw[->, >=latex] (Br2N4) -- (Br2N5);
\draw[->, >=latex] (Br2N5) -- (Br2N6);
\draw[->, >=latex] (Br2N6) -- (Br2N7);
\draw[->, >=latex] (Br2N7) -- (Br2N8);
\draw[->, >=latex] (Br2N8) -- (Br2N9);
\draw[->, >=latex] (Br2N9) -- (S12);

\draw[->, >=latex, gray, dashed] (Br3N1) -- (Br3N2);
\draw[->, >=latex, gray, dashed] (Br3N1p) -- (Br3N2);
\draw[->, >=latex, gray, dashed] (Br3N2) -- (Br3N3);
\draw[->, >=latex, gray, dashed] (Br3N3) -- (S5);

\draw[->, >=latex, gray, dashed] (Br4N1p1) -- (Br4N2);
\draw[->, >=latex, gray, dashed] (Br4N1p2) -- (Br4N2);
\draw[->, >=latex, gray, dashed] (Br4N1p3) -- (Br4N2);
\draw[->, >=latex, gray, dashed] (Br4N1p4) -- (Br4N2);
\draw[->, >=latex, gray, dashed] (Br4N2) -- (S7);

\draw[->, >=latex, gray, dashed] (Br5N1p1) -- (Br5N2);
\draw[->, >=latex, gray, dashed] (Br5N1p2) -- (Br5N2);

\draw[dashed] (Br5N2) -- (H2);
\draw[dashed] (S17) -- (H1);

\node[circle, draw, fill=black, inner sep=1pt, label=17] at (S17) {};
\node[circle, draw, fill=black, inner sep=1pt, label=below:16] at (S16) {};
\node[circle, draw, fill=black, inner sep=1pt, label=15] at (S15) {};
\node[circle, draw, fill=black, inner sep=1pt, label=14] at (S14) {};
\node[circle, draw, fill=black, inner sep=1pt, label=13] at (S13) {};
\node[circle, draw=purple, fill=purple, inner sep=1pt, label={[text=purple]above left:12}] at (S12) {};
\node[circle, draw=purple, fill=purple, inner sep=1pt, label={[text=purple]11}] at (S11) {};
\node[circle, draw=purple, fill=purple, inner sep=1pt, label={[text=purple]above left:10}] at (S10) {};
\node[circle, draw=purple, fill=purple, inner sep=1pt, label={[text=purple]9}] at (S9) {};
\node[circle, draw, fill=black, inner sep=1pt, label=8] at (S8) {};
\node[circle, draw, fill=black, inner sep=1pt, label=7] at (S7) {};
\node[circle, draw, fill=black, inner sep=1pt, label=6] at (S6) {};
\node[circle, draw, fill=black, inner sep=1pt, label=5] at (S5) {};
\node[circle, draw=red, fill=red, inner sep=1pt, label={[text=red]below:4}] at (S4) {};
\node[circle, draw=red, fill=red, inner sep=1pt, label={[text=red]3}] at (S3) {};
\node[circle, draw=red, fill=red, inner sep=1pt, label={[text=red]below:2}] at (S2) {};
\node[circle, draw=red, fill=red, inner sep=1pt, label={[text=red]below:1}] at (S1) {};

\node[circle, draw=red, fill=red, inner sep=1pt, label={[text=red]3}] at (Br1N3) {};
\node[circle, draw=red, fill=red, inner sep=1pt, label={[text=red]2}] at (Br1N2) {};
\node[circle, draw=red, fill=red, inner sep=1pt, label={[text=red]1}] at (Br1N1) {};

\node[circle, draw=red, fill=red, inner sep=1pt, label={[text=red]below:1}] at (Br2N1) {};
\node[circle, draw=red, fill=red, inner sep=1pt, label={[text=red]below:2}] at (Br2N2) {};
\node[circle, draw=red, fill=red, inner sep=1pt, label={[text=red]below:3}] at (Br2N3) {};
\node[circle, draw=red, fill=red, inner sep=1pt, label={[text=red]below:4}] at (Br2N4) {};
\node[circle, draw, fill=black, inner sep=1pt, label=below:5] at (Br2N5) {};
\node[circle, draw, fill=black, inner sep=1pt, label=below:6] at (Br2N6) {};
\node[circle, draw, fill=black, inner sep=1pt, label=below:7] at (Br2N7) {};
\node[circle, draw, fill=black, inner sep=1pt, label=below:8] at (Br2N8) {};
\node[circle, draw, fill=black, inner sep=1pt, label=9] at (Br2N9) {};

\node[circle, draw=gray, fill=gray, inner sep=1pt, label={[text=gray]below:3}] at (Br3N3) {};
\node[circle, draw=gray, fill=gray, inner sep=1pt, label={[text=gray]2}] at (Br3N2) {};
\node[circle, draw=gray, fill=gray, inner sep=1pt, label={[text=gray]below:1}] at (Br3N1) {};
\node[circle, draw=gray, fill=gray, inner sep=1pt, label={[text=gray]below:1}] at (Br3N1p) {};

\node[circle, draw=gray, fill=gray, inner sep=1pt, label={[text=gray]above left:2}] at (Br4N2) {};
\node[circle, draw=gray, fill=gray, inner sep=1pt, label={[text=gray]1}] at (Br4N1p1) {};
\node[circle, draw=gray, fill=gray, inner sep=1pt, label={[text=gray]below:1}] at (Br4N1p2) {};
\node[circle, draw=gray, fill=gray, inner sep=1pt, label={[text=gray]left:1}] at (Br4N1p3) {};
\node[circle, draw=gray, fill=gray, inner sep=1pt, label={[text=gray]left:1}] at (Br4N1p4) {};

\node[circle, draw=gray, fill=gray, inner sep=1pt, label={[text=gray]below:2}] at (Br5N2) {};
\node[circle, draw=gray, fill=gray, inner sep=1pt, label={[text=gray]1}] at (Br5N1p1) {};
\node[circle, draw=gray, fill=gray, inner sep=1pt, label={[text=gray]1}] at (Br5N1p2) {};

\draw[dashed] (-1,{1-sqrt(3)}) arc (-60:60:2cm) node[above] {$H$};
\end{tikzpicture}

\caption{Revert colors for vertices and edges in incomplete segments.\,\phantom{xxxxxxxxxxxxxxxxxx}\phantom{xxxxxxxxxx}}
\end{subfigure}

\caption{The coloring scheme applied to $G\backslash H$ with the choice of $\ell=4$. Vertices are annotated with their depths. Gray segments are presented with dashed edges, indicating that they are ignored during the subsequent sprinkling argument. Note that each red or purple segment has size at least $\ell$, and at any non-gray vertex, the colored segments that are (completely) contained in its (inclusive) upstream constitute at least $1/3$ of its regular (inclusive) upstream (see Lemma~\ref{Lemma_red_upstream_one_third}). Evidently, assuming $H\supseteq \Ctwo(G)\ne\emptyset$, the inclusion of any vertex into a cycle or a path between cycles during sprinkling will result in its entire downstream being included in the $2$-core.}\label{fig:coloring}
\end{figure}
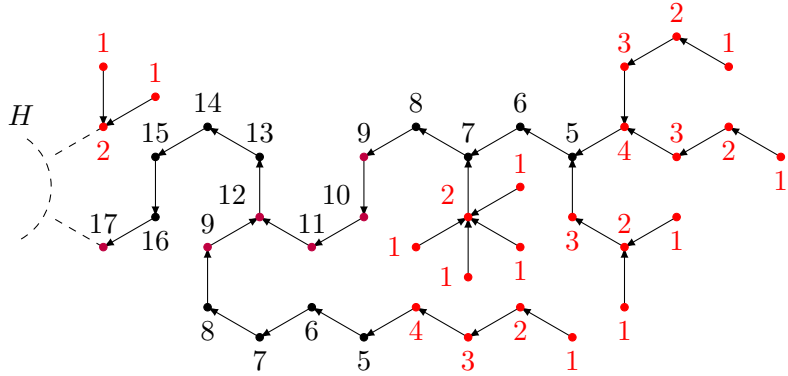
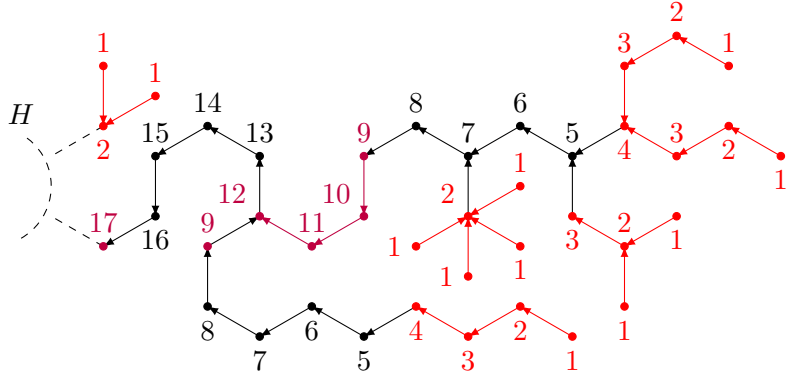
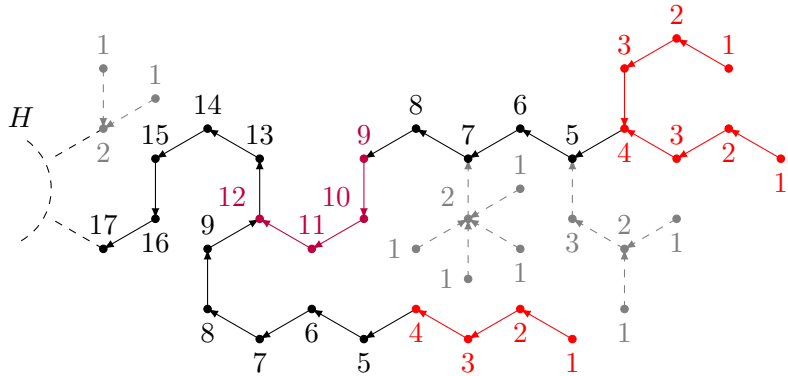
\break

\subsection{Proof details for Section~\ref{Sec_two_core_size_lower}}
\label{app:otherproofs}

This appendix is devoted to the proof details left out when we sketched the proof of the 
upper bound on the size of the giant $2$-core in Section~\ref{Sec_two_core_size_lower}.

We need some more notation.  Recall the color scheme from Section~\ref{Sec_two_core_size_lower}, in particular the notion of gray segments.  We did not include these in our sprinkling arguments  since they gray segments could potentially be small, and hence would have causes issues with our union bound, which used that all  segments have at least size $\ell$, a fact that {\it is} true for colored segments.

Define a segment to be \emph{regular} if it is not colored \gray. 
Recall from the proof of Proposition~\ref{Cor_exist_giant_two_core} the definition of upstream and downstream for a vertex, and for a vertex $u$, we call $\ius(u):= \{u\}\cup \us(u)$ the \emph{inclusive} upstream.
The regular inclusive upstream at a vertex $v\notin H$, denoted $\ius^\reg(v)$, refers to all regular vertices in $\ius(v)$; the colored inclusive upstream size at $v$, denoted $\ius^\col(v)$, is the union of colored segments of $G\backslash H$ that are (completely) contained in $\ius(v)$. %
The next lemma associates the size of the colored upstream with that of the regular upstream.  Before stating it, we note that any vertex $v$ in $G\setminus H$ with depth  $\ell$ or smaller cannot lie in $\Ctwoinfty(\cN_\ell^+(v))$, showing that any vertex in $\Ctwoell(G)\setminus H$ must have depth at least $\ell + 1$, and hence is either black or purple. See Figure~\ref{fig:coloring} for an illustration.

\begin{lemma}\label{Lemma_red_upstream_one_third}
For any regular vertices $v$, $|\ius^\col(v)| \ge \frac{1}{3}|\ius^\reg(v)|$.
\end{lemma}
\begin{remark}
Note that $\ius^\col(v)$ is obtained by removing from $\ius^\reg(v)$ black vertices and colored vertices in segments not fully contained in $\ius^\reg(v)$. The lemma bounds both effects.
\end{remark}

\begin{proof}
It is easy to see the claim when the regular upstream of $v$ is a single path of length $b = \ius^\reg(v)-1 \ge \ell$: the worst case occurs when $b = 3\ell-1$, in which case $\ius^\col(v) = \ell$ (due to the \red~segment at the leaf, as the next \purple~segment is incomplete) and the ratio is $\frac{\ell}{3\ell-1} \ge \frac{1}{3}$.

When the regular upstream of $v$ is not a single path, we may decompose it into stem paths, i.e., paths along vertices of consecutive depth, using the following process.
\begin{itemize}
    \item Initialize $U = \ius^\reg(v)$, inheriting the depth numbers and colors from the original graph.
    \item While $U$ is non-empty, iteratively find the vertex $u$ with maximum depth (as inherited from $G$) in $U$, breaking tie arbitrarily if necessary.
    \begin{itemize}
        \item Terminate if $u$ is \red: all remaining vertices in $H$ must be \red.
        \item Find a stem path to $u$, using any one if multiple exist.\footnote{This is always possible: If $u$ has not been removed, neither must be any of its direct upstream $w$, since $w$ has a strictly smaller depth than $u$ and cannot have been removed as part of a previous path (which must also go through $u$, the unique direct downstream of $w$.} Extract and remove it from $U$. The length of this stem path is equal to $\depth(u)-1$, which is at least $\ell$ since $u$ is not red.
    \end{itemize}
\end{itemize}
The process above will decompose the upstream of $v$ into multiple stem upstreams of length at least $\ell$ and some extra red vertices. The $1/3$ lower bound for colored ratio holds for each stem upstream (where we only count relatively complete colored segments in each path), and assembling the paths back into the upstream of $v$ will only increase the number of relatively complete colored segments.%
\end{proof}

\begin{proof}[Proof of Lemma~\ref{Lemma_twocore_size_lower_reduction}]
    By continuity of $\zetatwoinfty$ for $p\neq p_c$ (which we established in Appendix~\ref{appendix_continuity}), we may choose $p'\in(p_c,p)$ such that $\zetatwoinfty(p') \ge \zetatwoinfty(p)-\frac{\eps}{4}$. 
    For $\ell$ sufficiently large, Corollary~\ref{Lemma_Ztwoinfty_k_of_C1_at_p_close_to_limit} ensures
    \begin{equation}
        \frac{1}{n}|\Ctwoell(\Cmax(G_n(p')))| \ge \zetatwoinfty(p')-\frac{\eps}{4} \ge \zetatwoinfty(p)-\frac{\eps}{2}\qquad\text{ whp},
    \end{equation}
    from which the reduction from \eqref{Eqn_prop_twocore_size_lower_main} to \eqref{Eqn_prop_twocore_size_lower_proof_STP} follows.

    Assume that the event in \eqref{Eqn_prop_twocore_size_lower_proof_STP} fails; that is, there are more than $\frac{\eps n}{2}$ vertices in $\Ctwoell(\Cmax(G_n(p')))$ that fail to join $\Ctwo(\Cmax(G_n(p)))$. We claim that at least $ \frac{3\eps n}{8}$ of these vertices are in $\Ctwoell(\Cmax(G_n(p'))) \backslash H$ and hence are again colored black or purple.  To see this, we consider two cases: (1) if $H = \Ctwo(\Cmax(G_n(p')))$, then this follows from $\Ctwo(\Cmax(G_n(p)))\supseteq \Ctwo(\Cmax(G_n(p')))=H$; (2) if $H \supsetneq \Ctwo(\Cmax(G_n(p')))$, then $|H|=\floor{\frac{\eps n}{8}}$ per our construction, and at least $\frac{\eps n}{2} - |H| \ge \frac{3\eps n}{8}$ of these vertices are in $\Ctwoell(\Cmax(G_n(p'))) \backslash H$, as claimed.

Next, we claim that if a vertex $v\in\Ctwoell(\Cmax(G_n(p')))\backslash H$ stays outside $\Ctwo(G_n(p))$, then so must its entire upstream. To see this, we show the contrapositive by considering $u\in\us(v)\cap\Ctwo(G_n(p))$ with a case-by-case discussion:
\begin{itemize}
    \item If $u$ is in a cycle $\Gamma\subseteq G_n(p)$, then there is a (simple) path from $u\in\Gamma$ to some cycle in $\Ctwo(\Cmax(G_n(p')))$ that traverses the entire downstream of $u$. Then $v$ is on a path between cycles (if not directly in the cycle $\Gamma$);
    \item Otherwise, $u$ sits on a path between two cycles $\Gamma_1,\Gamma_2\subseteq G_n(p)$. Hence, there are two disjoint paths from $u$ to $\Gamma_1$ and $\Gamma_2$, and at least one of them does not overlap with the downstream of $u$ to $v$ (otherwise, $u$ is in a cycle). Concatenating this path with the entire downstream of $u$ gives a simple path from $\Ctwo(\Cmax(G_n(p')))$, which contains a cycle, to $\Gamma_1$ or $\Gamma_2$, again certifying $v$ as on a path between cycles.
\end{itemize} %

By this observation, we can identify all black and purple vertices $v_1,\ldots,v_J$ outside $\Ctwo(G_n(p))$ that are maximally downstream (i.e., no other black or purple vertices outside $\Ctwo(G_n(p))$ are in the downstream of them), with
their downstream vertices all merged into $\Ctwo(G_n(p))$ while their inclusive upstreams remaining outside. In particular, any black or purple vertex not in $\Ctwo(G_n(p))$ must be in the inclusive upstream of $v_j$ for some $j\in[J]$; there will be at least $\frac{3\eps n}{8}$ of them when the event in \eqref{Eqn_prop_twocore_size_lower_proof_STP} fails. Thus, we must have
$\sum_{j=1}^J |\ius^\reg(v_j)| \ge \frac{3\eps n}{8}$.
By Lemma~\ref{Lemma_red_upstream_one_third}, $\sum_{j=1}^J |\ius^\col(v_j)| \ge \frac{\eps n}{8}$.
In other words, there exist a set of colored segments in $\Cmax(G_n(p'))$ of combined size $\frac{\eps n}{8}$ or more that remain fully outside $\Ctwo(G_n(p))$, as we claimed. %
\end{proof}

\begin{proof}[Proof of Proposition~\ref{Prop_twocore_size_lower}]
We assume without loss of generality that $p>p_c$ and $\zetatwoinfty(p) > \eps$.
As in Lemma~\ref{Lemma_twocore_size_lower_reduction}, we choose $p'<p$ sufficiently close to $p$ such that $\zetatwoinfty(p') \ge \zetatwoinfty(p)-\frac{\eps}{4}$, and choose $\ell$ sufficiently large such that
$\frac{1}{n}|\Ctwoell(\Cmax(G_n(p')))| \ge \zetatwoinfty(p')-\frac{\eps}{4}$
and $\ell > \max\big\{ \frac{4}{\delta}, \frac{8}{\beta^L \delta} \big\}$.
By Lemma~\ref{Lemma_twocore_size_lower_sprinkling}, 
the probability of having a set $S$ of colored segments with combined size at least $\frac{\eps n}{8}$ that have empty intersection with $\Ctwo(G_n(p))$ is at most
\begin{equation}
    2^{n/\ell} \cdot (1-\beta^L)^{\frac{\delta n}{2} - \frac{n}{\ell}} \le e^{\frac{n}{\ell} - \beta^L\cdot\frac{\delta n}{4}} \le e^{-\beta^L\delta n/8} \to 0 \qquad\text{ as }n\to\infty.
\end{equation}
Thus, by Lemma~\ref{Lemma_twocore_size_lower_reduction}, \eqref{Eqn_prop_twocore_size_lower_proof_STP} holds whp and our proof is finished.
\end{proof}

\section{Robustness of Algorithm~\ref{Alg_two_core}: Proof of Theorem~\ref{Thm_main_results_robust}}\label{Appendix_alg_robust}

\begin{proof}[Proof of Theorem~\ref{Thm_main_results_robust}]
Let $\mu$ denote the local limit of the sequence $\{G_n\}_{n\in\N}$ as usual.
First, by Theorems~\ref{Thm_main_two_core_max} and~\ref{Thm_main_two_core_entire}, we know that for all $p\ne p_c(\mu)$,
\begin{equation}
    \frac{1}{n}|\Ctwo(G_n(p))| \overset{p}{\to} \zetatwo(p;\mu) \qquad\text{ and }\qquad \frac{1}{n}|\Ctwomax(G_n(p))| \overset{p}{\to} \zetatwoinfty(p;\mu).
\end{equation}
Thus, it suffices to have $I^{(2)}$ (resp. $I^{(2,\infty)}$) in the $\frac{\eps}{2}$-neighborhood of $\zetatwo(p)$  (resp. $\zetatwoinfty(p)$) with probability $1-\frac{\eps}{2}$. For now, we focus on $I^{(2,\infty)}$ as an estimate of $\zetatwoinfty(p)$; the analysis for $I^{(2)}$ is analogous.

Algorithm~\ref{Alg_two_core} realizes $T$ independently chosen neighborhood in the percolated graph, and in particular, $\{I^{(2,\infty)}_t\}_{t\in[T]}$ are i.i.d. Bernoulli random variables (conditional on the input graph $G_n$ if it is random). For any $\ell\in\N$ (to be specified later), we can decompose $\E[I_t^{(2,\infty)} | G_n(p)] = \Pp(v_t \in C^{(2,R_t)}(G_n(p)) | G_n(p))$ into
\begin{equation}\label{Eqn_analysis_alg_1_single_round_Expect}
    \Pp(v_t \in C^{(2,R_t)}(G_n(p)), R_t \ge \ell | G_n(p)) + \Pp(v_t \in C^{(2,R_t)}(G_n(p)), R_t < \ell | G_n(p)).
\end{equation}
For the first term on the right hand side in \eqref{Eqn_analysis_alg_1_single_round_Expect}, we have
\begin{equation}
    \Pp(v_t \in C^{(2,R_t)}(G_n(p)), R_t \ge \ell | G_n(p)) \le \Pp(v_t \in \Ctwoell(G_n(p)) | G_n(p)).
\end{equation}
By Lemma~\ref{Lemma_3_1_local_event_func_in_prob}, $\Pp(v_t \in C^{(2,\ell)}(G_n(p)) | G_n(p)) \overset{p}{\to} \zetatwoell(p)$,
and since $\lim_{\ell\to\infty}\zetatwoell(p) = \zetatwoinfty(p)$, by choosing $\ell$ sufficiently large, we can guarantee that
\begin{equation}
    \Pp\big(v_t \in C^{(2,R_t)}(G_n(p)), R_t \ge \ell \big| G_n(p)\big) \le \zetatwoinfty(p) + \frac{\eps}{8} \qquad\text{ whp}.
\end{equation}
For the second term, note that
\begin{equation}
    \Pp(v_t \in C^{(2,R_t)}(G_n(p)),\, R_t < \ell | G_n(p)) \le \Pp(R_t < \ell | G_n(p)) \le \Pp(|\cN_{\ell}(v_t;G_n(p))| > K | G_n(p)).
\end{equation}
By Lemma~\ref{Lemma_bounded_nbrh_size}, for $K$ sufficiently large, this quantity is at most $\eps/8$ whp. Thus,
\begin{equation}\label{Eqn_proof_alg_robust_E_It_upper}
    \E\big[I_t^{(2,\infty)} \big| G_n(p)\big] \le \zetatwoinfty(p) + \frac{\eps}{4} \qquad\text{ whp}.
\end{equation}

On the other hand, since $R_t \le K$ and $C^{(2,R_t)}(G_n(p)) \supseteq C^{(2,K)}(G_n(p))$, we have
\begin{equation}
    \Pp(v_t \in C^{(2,R_t)}(G_n(p)) | G_n(p)) \ge \Pp(v_t \in C^{(2,K)}(G_n(p)) | G_n(p)).
\end{equation}
With analogous reasoning as before, we can guarantee
\begin{equation}\label{Eqn_proof_alg_robust_E_It_lower}
    \E\big[I_t^{(2,\infty)} \big| G_n(p)\big] = \Pp(v_t \in C^{(2,R_t)}(G_n(p)) | G_n(p)) \ge \zetatwoinfty(p) - \frac{\eps}{4} \qquad\text{ whp}
\end{equation}
by choosing $K$ sufficiently large. Combining \eqref{Eqn_proof_alg_robust_E_It_upper} with \eqref{Eqn_proof_alg_robust_E_It_lower} suggests $\big|\E\big[I_t^{(2,\infty)} \big| G_n(p)\big] - \zetatwoinfty(p)\big| \le \frac{\eps}{4}$ whp.

The rest is straightforward. Since $I^{(2,\infty)}$ is the average of $T$ i.i.d. Bernoulli random variables $\big\{I_t^{(2,\infty)}\big\}_{t\in[T]}$ (conditional on $G_n(p)$), by the standard Hoeffding's inequality,
\begin{equation}
    \Pp\Big(\big|I^{(2,\infty)} - \E\big[I_t^{(2,\infty)} \big| G_n(p)\big] \big| \ge \frac{\eps}{4} \Big| G_n(p)\Big) \le \frac{\eps}{4}
\end{equation}
for $T = O\big(\frac{1}{\eps^2}\log\frac{1}{\eps}\big)$, and marginalizing over $G_n$ preserves the bound. Thus, we finally have $\big|I^{(2,\infty)} - \zetatwoinfty(p)\big| \le \eps/2$ with probability at least $1-\eps/2$. Combining this with the in-probability convergence of $\frac{1}{n}|\Ctwomax(G_n(p))|$ to $\zetatwoinfty(p)$ concludes the proof.
\end{proof}

\newpage

\bibliographystyle{plain}
\bibliography{ref}

\end{document}